\documentclass[reprint,amsmath,amssymb,aps,pra]{revtex4-2}

\usepackage{amsmath}
\usepackage{graphicx}
\usepackage{dcolumn}
\usepackage{bm}
\usepackage{braket}
\usepackage{dsfont}
\usepackage[caption=false]{subfig}
\usepackage{hyperref}
\hypersetup{
    colorlinks = true,
    linkcolor = blue,
    filecolor = blue,
    citecolor = blue,      
    urlcolor = blue,
}

\begin{document}
	
	\title{Quantum State Transfer Between Distant Optomechanical Interfaces via Shortcut to Adiabaticity}
	
	\author{Hanzhe Xi}
	\author{Pei Pei}
	 \email{peipei@dlmu.edu.cn}
	\affiliation{College of Science, Dalian Maritime University, Dalian 116026, China}
	
	\date{\today}
	
	\begin{abstract}
		We propose a protocol to realize fast high-fidelity quantum state transfer between distant optomechanical interfaces connected by a continuum waveguide. The scheme consists of three steps: two accelerating adiabatic processes joined by a population conversion process. In comparison to the traditional adiabatic technique, our method reaches a higher transfer fidelity with a shorter time. Numerical results show that the fidelity of this transfer scheme in the dissipative system mainly depends on the protocol speed and the coupling strength of the waveguide and cavities. Assisted by inverting the pulse sequence, a bidirectional transfer can be implemented, indicating the potential to build a quantum network.
	\end{abstract}
	
	\maketitle
	
	\section{Introduction}
	The adiabatic quantum process is an efficient and robust approach that can be utilized in various quantum operations. One of the most well-known techniques is stimulated Raman adiabatic passage (STIRAP) \cite{vitanov2017stimulated,kuhn2019roadmap}. It transfers population between source quantum state and target quantum state by coupling them with two radiation fields via an intermediate state. In past decades, STIRAP has been applied in wide range area including atomic and molecular physics (such as atom optics \cite{kulin1997coherent,theuer1998atomic}, cavity quantum electrodynamics \cite{nolleke2013efficient}, ultracold molecules \cite{takekoshi2014ultracold}), quantum information (such as single- and two-qubit gates \cite{beterov2013quantum}, entangled-state preparation \cite{noguchi2012generation}) and solid-state physics (such as nitrogen-vacancy centers \cite{golter2016optomechanical}, superconducting circuits \cite{kumar2016stimulated,xu2016coherent}, semiconductor quantum dots and wells \cite{simon2011robust,tomaino2012terahertz}). Although STIRAP is robust against small variations of laser intensity, pulse timing, pulse shape and some other experimental parameters, it is necessarily slow so that it is vulnerable to dissipation or fluctuations. Thus finding ways to accelerate adiabatic evolution arouses great interest, and such techniques are called “shortcuts to adiabaticity” (STA). The shortcuts rely on specific time dependences of the control parameters or the auxiliary couplings with respect to the reference Hamiltonian \cite{guery2019shortcuts}. So far, multiple STA methods (including counterdiabatic driving \cite{baksic2016speeding,sels2017minimizing,petiziol2018fast}, invariant-based inverse engineering \cite{kiely2014inhibiting,kiran2021invariant,han2021realization} and fast forward approach \cite{masuda2010fast,torrontegui2012shortcuts,patra2021semiclassical}) have been proposed and applied to various quantum systems both theoretically and experimentally \cite{faure2019shortcut,vepsalainen2019superadiabatic,yan2019experimental,qiu2021experimental,chen2021shortcuts}. One of the most important applications of STA is quantum state transfer \cite{huang2018quantum,mortensen2018fast,petiziol2020optimized,zhang2021population}. The transfer fidelity is mainly affected by decoherence due to the long evolution time, while STA can speed up this adiabatic process and remedy this vulnerability. 
	
	In this work, we propose a quantum state transfer scheme via shortcut to adiabaticity for the system with two distant optomechanical interfaces connected by a continuum waveguide (optical fiber). The optomechanical interface is composed of a superconducting resonator (SR), an optical cavity (OC) and a nanomechanical resonator (NAMR) as an intermediate level. Such system has been provided as a powerful medium for high fidelity quantum state conversion in Ref. \cite{wang2012using,tian2012adiabatic}. At the very beginning, the state is in the superconducting resonator of the local part (Node A). We utilize a modified superadiabatic transitionless driving (SATD$+\kappa$) to transfer the initial quantum state to the continuum waveguide \cite{baksic2017shortcuts}. After the population transfer to the waveguide, we use a precise time-control coupling to make conversion between the waveguide and the cavity of the remote part (Node B). Finally, a SATD approach \cite{baksic2016speeding} is applied so that quantum state transfers to the superconducting resonator in node B. This scheme, compared with the similar scheme in which the last step is a STIRAP instead of a SATD, gives a higher transfer fidelity and costs less operation time. We also take dissipation into account, and the transfer fidelity highly depends on the protocol speed in this situation.
	
	\begin{figure}[htp!]
		\subfloat{
			\centering
			\includegraphics[width=0.45\textwidth]{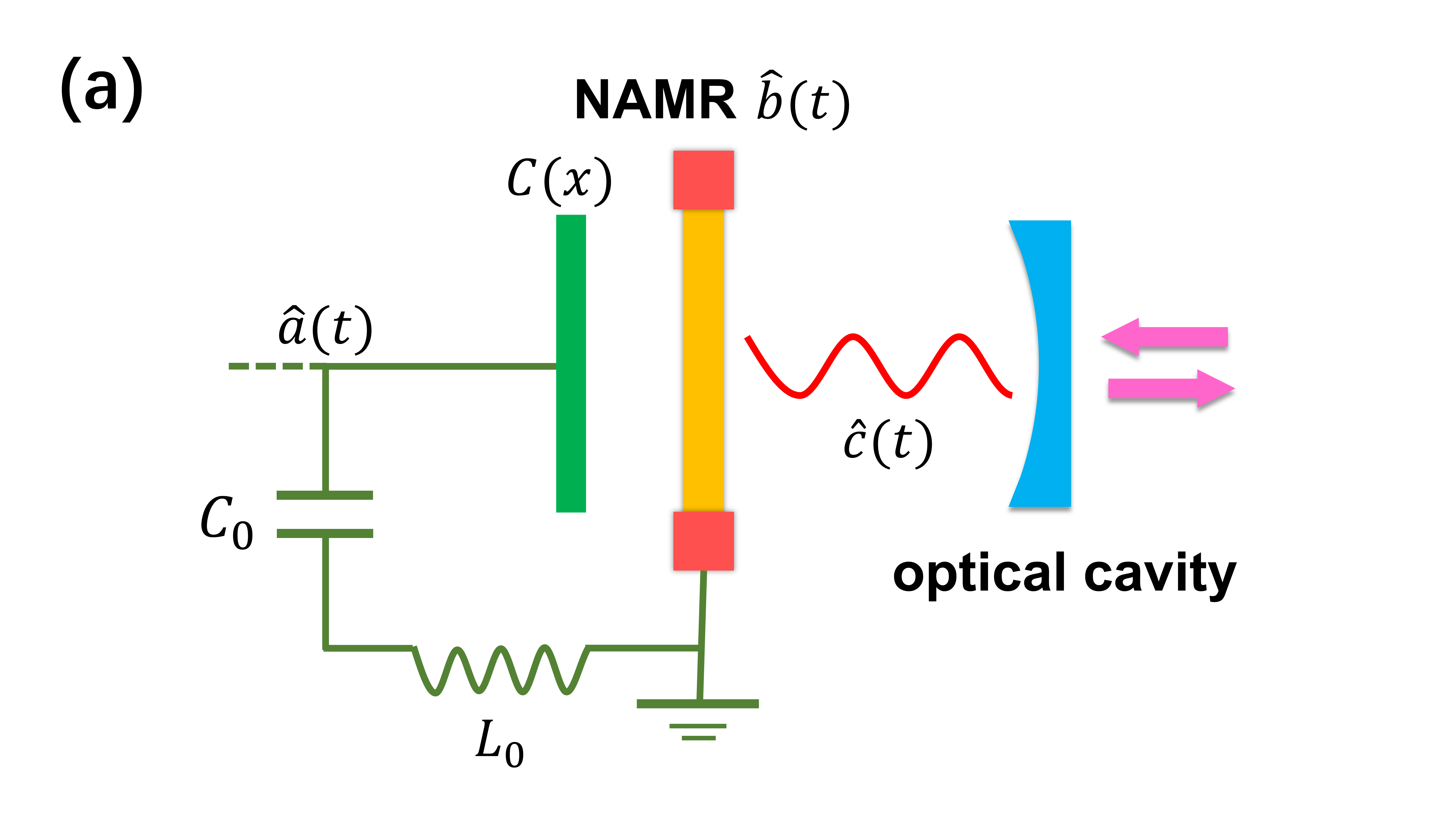}
			\label{fig1a}
		}
		\\
		\subfloat{
			\centering
			\includegraphics[width=0.45\textwidth]{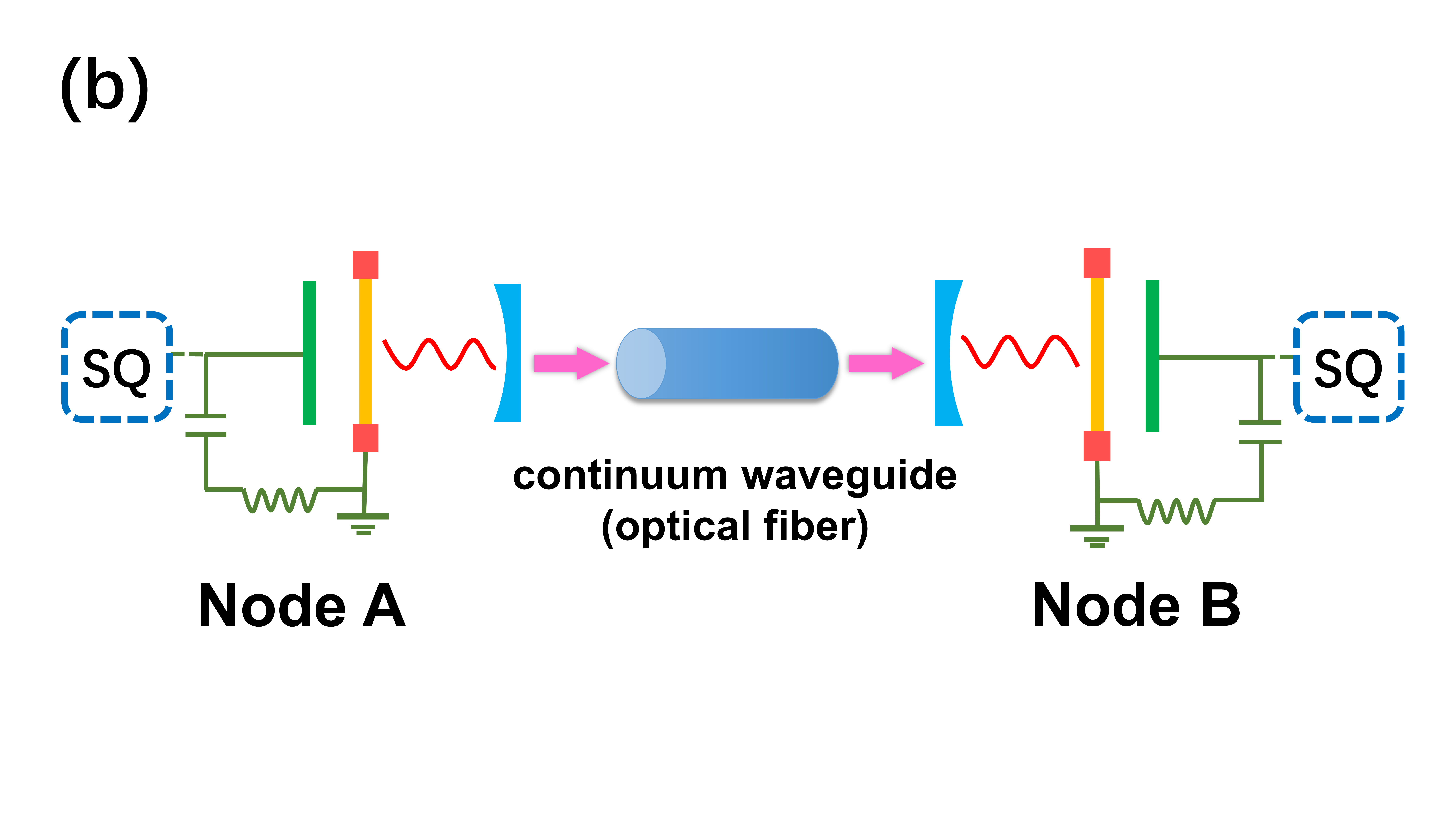}
			\label{fig1b}
		}
		\caption{\label{fig1}(a): Schematic diagram of the optomechanical quantum interface. nanomechanical resonator mode $\hat{b}$ couples to the mode $\hat{a}$ of superconducting resonator and the mode $\hat{c}$ of optical cavity. (b): Schematic of the fast high-fidelity state transfer protocol. The whole scheme contains two optimechanical interfaces (node A and node B) and a continuum waveguide. Distant nodes are connected by the waveguide . (Each node can be attached to a superconducting qubit.)
		}
		
	\end{figure}

	\section{System}
	As shown in Fig. \ref{fig1}, the whole protocol contains two interfaces and an optical continuum waveguide. Each interface contains a superconducting resonator, an optical cavity and a nanomechanical resonator that connects SR and cavity. Distant interfaces are joined by the optical fiber. Our scheme consists of two parts: local part and remote part, and we will introduce them respectively.
	
	\subsection{Local Part Operation}
	In the beginning, the target quantum state stays in the SR of the optomechanical interface in node A and the state is first transferred to the OC and leaks to the waveguide simultaneously by the SATD+$\kappa$ scheme. Thus, the local part system Hamiltonian in the interaction picture, following the standard linearization procedure and rotating wave approximation (RWA) \cite{aspelmeyer2014cavity,yin2015quantum}, takes the form $ H_A =  H_l(t) + H_{\rm int}(t) + H_{\rm WG} $, with $\hbar = 1$ (here and hereafter) 
	\begin{eqnarray}
		H_l(t) &=&  G_{1l}(t) \ket{A_l} \bra{B_l} + G_{2l}(t)\ket{C_l}\bra{B_l} + {\rm H.c.} , \nonumber\\
		H_{\rm int}(t) &=& \int^{\omega_{max}/2}_{-\omega_{max}/2} d\omega \, G_{3l}(\omega,t)\left[ \ket{C_l}\bra{D_\omega} + \ket{D_\omega}\bra{C_l} \right], \nonumber\\
		H_{\rm WG} &=& \int^{\omega_{max}/2}_{-\omega_{max}/2} d\omega \,\omega\ket{D_\omega}\bra{D_\omega},
	\end{eqnarray}
	$\ket{A_l}$, $\ket{B_l}$ and $\ket{C_l}$ are the excitation states of SR, NAMR and OC in node A respectively , with the state $\ket{C_l}$ additionally coupled to a continuum waveguide. $\ket{D_\omega}$ is the excitation state in the continuum waveguide at frequency $\omega$. $G_{1l}(t)$, $G_{2l}(t)$ and $G_{3l}(\omega,t)$ are three time-dependent couplings that can be tuned independently. 
	
	We consider the continuum waveguide with a finite bandwidth $\omega_{max}$, and the amplitude of the interaction between the state $\ket{C_l}$ and the waveguide state $\ket{D_\omega}$ is frequency independent $\left[G_{3l}(\omega,t)=G_{3l}(t), \forall |\omega| \leq \omega_{max}/2\right]$. As the waveguide bandwidth is much greater than any other frequency scales, we can take $\omega_{max} \rightarrow \infty$. And we will take the Markovian regime throughout the whole system. 
	
	The solution of Schr\"odinger equation with $H_A$ takes the form
	\begin{eqnarray}\label{amplitude}
		\ket{\psi (t)} &=& u_A(t)\ket{A_l} + u_B(t)\ket{B_l} + u_C(t)\ket{C_l}  \nonumber \\
		&& + \int_{-\infty}^{+\infty}d\omega \, u_{\rm WG}(\omega,t)\ket{D_\omega}.
	\end{eqnarray}
	We can formally solve the Schr\"odinger equation for the waveguide amplitude $u_{\rm WG}(\omega,t)$, and use it to simplify the Schr\"odinger equation for the remaining amplitudes. Thus we can rewrite the Hamiltonian $H_A$ as an effective non-Hermitian Hamiltonian \cite{baksic2017shortcuts}
	\begin{eqnarray}
	    H_{A1} = H_l(t) -  i\left(\pi|G_{3l}(t)|^2\right)\ket{C_l}\bra{C_l},
	\end{eqnarray}
	where the Hermitian part $H_l(t)$ has a set of adiabatic eigenstates that are given by
	\begin{eqnarray}\label{ADstates}
		\ket{+} &=& \frac{1}{\sqrt{2}}\left(\sin\theta(t)\ket{A_l} + \ket{B_l} + \cos\theta(t)\ket{C_l} \right), \nonumber \\
		\ket{\rm dk} &=& - \cos\theta(t)\ket{A_l} + \sin\theta(t)\ket{C_l}, \nonumber \\
		\ket{-} &=& \frac{1}{\sqrt{2}}\left(\sin\theta(t)\ket{A_l} - \ket{B_l} + \cos\theta(t)\ket{C_l}\right),
	\end{eqnarray}
	where the control fields are set as $G_{1l}(t) = g_0\sin\theta_1(t)$ and $G_{2l}(t) = g_0 \cos\theta_1(t)$ with the mixing angle $\theta_1(t)$. 
	
	Firstly we transform $H_{A1}$ to the adiabatic frame via a time-dependent unitary operator $ U_{\rm ad} = \sum_{k=\pm,\rm dk}\ket{k(t)}\bra{k} $, so that the effective non-Hermitian Hamiltonian turns to be
	\begin{eqnarray}
		H_{A1,\rm ad}(t) = U_{\rm ad}^\dagger(t) H_{A1}(t) U_{\rm ad}(t) - iU_{\rm ad}^\dagger(t)\frac{d}{dt}U_{\rm ad}(t). \nonumber\\
	\end{eqnarray}
	
	Now we introduce dressed states $ \ket{\tilde{k}(t)} \equiv V(t)\ket{k(t)} (k =\pm,\rm dk) $ defined by a unitary operator $V(t)$. The dressing operator $V(t)$ satisfies the condition that at the initial and final protocol time the dressed states coincide with the adiabatic states i.e. $\left[ V(t_i)=V(t_f)=\mathds{1} \right]$. We also modify the couplings $G_{1l}(t)$ and $G_{2l}(t)$ by a correction Hamiltonian $H_{cl}(t)$, so that there are no transitions between the dressed dark state and other dressed adiabatic states during the dynamics,
	\begin{eqnarray}\label{transition condition}
		\bra{\tilde{+}}H_{A1,\rm ds}(t)\ket{\tilde{\rm dk}}=\bra{\tilde{-}}H_{A1,\rm ds}(t)\ket{\tilde{\rm dk}}=0.
	\end{eqnarray}
	Thus the effective non-Hermitian Hamiltonian can be transformed into the dressed frame 
	\begin{eqnarray}
		H_{A1,\rm ds}(t) &=& V^\dagger(t)\left[H_{A1,\rm ad}(t) + U^\dagger_{\rm ad}(t)H_{cl}(t)U_{\rm ad}(t)\right]V(t) \nonumber \\
		&& -iV^\dagger(t)\frac{d}{dt}V(t).
	\end{eqnarray} 

	In order to satisfy the above constraints, the dressing operator is taken as
	\begin{eqnarray}\label{V(t)}
		V(t) = \exp \left[i \mu_1(t) \left(\frac{\ket{+}-\ket{-}}{\sqrt{2}} \bra{\rm dk} + {\rm H.c.}\right)\right],
	\end{eqnarray}
	Where $\mu_1(t)$ parametrizes the dressing strength at time $t$ and must tend to zero at the initial and the end time of the protocol. And the added correction Hamiltonian $H_{cl}(t)$ is parametrized via $g_{xl}(t)$ and $g_{zl}(t)$
	\begin{eqnarray}\label{correction H}
		H_{cl}(t) &=& U_{\rm ad} \bigg[ g_{xl}(t)\left(\frac{\ket{+}-\ket{-}}{\sqrt{2}}\bra{\rm dk} + {\rm H.c.}\right)  \nonumber\\
		&& + g_{zl}(t)\left(\ket{+}\bra{+}-\ket{-}\bra{-}\right) \bigg] U_{\rm ad}^\dagger(t).
	\end{eqnarray}
	Therefore the corrected couplings $G_{1l}(t)$ and $G_{2l}(t)$ take the form, 
	\begin{eqnarray} \label{Gl}
		G_{1lc}(t) &=& G_{1l}(t) - g_{xl}(t)\cos\theta_1(t) + g_{zl}(t)\sin\theta_1(t), \nonumber \\
		G_{2lc}(t) &=& G_{2l}(t) + g_{xl}(t)\sin\theta_1(t) + g_{zl}(t)\cos\theta_1(t). 
	\end{eqnarray}
	
	Using these definitions and constraints, we can obtain the expressions of $g_{xl}(t)$ and $g_{zl}(t)$
	\begin{eqnarray}
		g_{xl}(t) &=& -\dot{\mu_1} + \frac{2\pi [G_{3l}(t)]^2}{4}\sin^2\left[\theta_1(t)\right]\sin\left[2\mu_1(t)\right], \label{gx} \\ 
		g_{zl}(t) &=& \frac{1}{\tan\mu_1(t)}\left[\dot{\theta}_1(t) + \frac{2\pi [G_{3l}(t)]^2}{4}\sin[2\theta_1(t)]\right] - g_0. \nonumber \\ \label{gz}
	\end{eqnarray}
	Since $G_{1l}(t)$ and $G_{2l}(t)$ are controllable, we can get the simplest nontrivial correction by choosing $g_{zl}(t)=0$. Thus we can easily obtain the dressing strength $\mu_1(t)$ by Eq. \ref{gz} 
	\begin{eqnarray}
		\mu_1(t) = \arctan\left[\frac{\dot{\theta}_1(t) + (2\pi [G_{3l}(t)]^2/4)\sin[2\theta_1(t)]}{g_0}\right].
	\end{eqnarray}
	With $\mu_1(t)$ determined, the modified pulses are immediately given by Eq.\ref{Gl} and Eq.\ref{gx}.
	
	Using this approach, the initial state of the SR in node A will transfer to the cavity through a NAMR and eventually leak to the waveguide. And the fidelity $F_l(t)$ of the SATD+$\kappa$ operation at time $t$, can be defined as the population in the waveguide
	\begin{eqnarray}\label{fl}
		F_l(t)=\int d\omega|u_{\rm WG}(\omega,t)|^2=\int_{t_i}^{t}d\tau\: 2\pi [G_{3l}(\tau)]^2|u_C(\tau) |^2. \nonumber \\
	\end{eqnarray}
	
	\subsection{Remote Part Operation}
	After the operation in node A, we start to transfer the population in the waveguide to the SR in node B. Noted that the sequence with the cavity-waveguide coupling in the SATD+$\kappa$ scheme makes the evolution in node A irreversible, and simply reverse the pulse sequence in the SATD+$\kappa$ scheme can not realize the inverse process. So we propose a two-step scheme to transfer the population from the waveguide to the SR in node B. First, a population conversion between the waveguide and the cavity in node B is implemented. After the population completely moves to the cavity, we use the SATD scheme to transfer the population to the SR in node B. For the population conversion process, we have to turn off the coupling between the waveguide and the cavity in node A and only keep the coupling between the waveguide and the cavity in node B in order to avoid the population moving back to node A. Thus the Hamiltonian of this process in the rotating frame after RWA takes the form
	\begin{eqnarray}
		H_{\rm con}(t) = \int^{\omega_{max}/2}_{-\omega_{max}/2} d\omega \, G_{3r}(\omega,t)\left[ \ket{C_r}\bra{D_\omega} + \ket{D_\omega}\bra{C_r} \right], \nonumber \\
	\end{eqnarray}
	Here $\ket{C_r}$ is the excitation state of the cavity in node B and $G_{3r}(\omega,t)$ is a tunable time-dependent coupling. We take the same consideration which is introduced in node A part: the amplitude of the interaction between $\ket{D_\omega}$ and $\ket{C_r}$ is frequency independent $\left[G_{3r}(\omega,t)=G_{3r}(t), \forall |\omega| \leq \omega_{max}/2\right]$; And we take $\omega_{max} \rightarrow \infty$. The states in the waveguide will completely transfer to the cavity in node B when the coupling turns on for a proper time interval $t_c = \pi/ (2|G_3r|)$. 
	
	After the conversion process, we turn off the waveguide-cavity coupling and then apply a SATD approach to transfer the quantum state from the cavity to the SR in node B. The Hamiltonian of the three-level interface system in the interaction picture, after the the standard linearization procedure and RWA, is given by
	\begin{eqnarray}
		H_B(t) = G_{1r}(t)\ket{C_r}\bra{B_r} + G_{2r}(t)\ket{A_r}\bra{B_r} + {\rm H.c.} \  , \nonumber\\
	\end{eqnarray}
	Where $\ket{A_r}$ and $\ket{B_r}$ are excitation states of SR and NAMR. $G_{1r}$ and $G_{2r}$ are two tunable time-dependent couplings. The adiabatic states of $H_B(t)$ have the same form in the Eq. \ref{ADstates}. Thus in the adiabatic frame, the Hamiltonian of three-level interface system  becomes 
	\begin{eqnarray}
		H_{B,\rm ad}(t) = U_{\rm ad}^\dagger(t) H_{B}(t) U_{\rm ad}(t) - iU_{\rm ad}^\dagger(t)\frac{d}{dt}U_{\rm ad}(t). \nonumber\\
	\end{eqnarray}

	The dressing operator parametrized by $\mu_2(t)$, takes the same form with Eq. \ref{V(t)}. In the dressed frame, the transformed Hamiltonian is given by
	\begin{eqnarray}
		H_{B,\rm ds}(t) &=& V^\dagger(t)\left[H_{B,\rm ad}(t) + U^\dagger_{\rm ad}(t)H_{cr}(t)U_{\rm ad}(t)\right]V(t) \nonumber \\
		&& -iV^\dagger(t)\frac{d}{dt}V(t).
	\end{eqnarray}
	
	With the same correction Hamiltonian form mentioned in Eq. \ref{correction H}, the modified couplings $G_{1rc}(t)$ and $G_{2rc}(t)$ are read as follows
	\begin{eqnarray}\label{Grc}
		G_{1rc}(t) &=& G_{1r}(t) - g_{xr}(t)\cos\theta_2(t) + g_{zr}(t)\sin\theta_2(t), \nonumber \\
		G_{2rc}(t) &=& G_{2r}(t) + g_{xr}(t)\sin\theta_2(t) + g_{zr}(t)\cos\theta_2(t). \nonumber \\
	\end{eqnarray}
	
	In order to satisfy the condition in Eq. \ref{transition condition} that there are no transitions between the dressed dark state $\ket{\tilde{\rm dk}}$ and the other dressed adiabatic states $\ket{\tilde{\pm}}$ during the dynamics, $g_{xr}(t)$ and $g_{zr}(t)$ are given by
	\begin{eqnarray}
		g_{xr}(t) &=& -\dot{\mu}_2(t), \label{gxr} \\ 
		g_{zr}(t) &=& \frac{\dot{\theta}_2(t)}{\tan\mu_2(t)} - g_0.  \label{gzr}
	\end{eqnarray}	
	Here we choose the simplest nontrivial choice $\left(g_{zr}=0\right)$ \cite{baksic2016speeding}. Thus the parameter function $\mu_2(t)$ takes the form 
	\begin{eqnarray}
		\mu_2(t) = \arctan\left[\frac{\dot{\theta}_2(t)}{g_0}\right].
	\end{eqnarray}
	
	With $\mu_2(t)$ in hands, we can obtain the modified couplings by Eq. \ref{Grc}. The final fidelity of the three-step operation can be defined as the population in the SR of node B 
	\begin{eqnarray}
		F_{e} = |u_{c'}(t_e)|^2.
	\end{eqnarray} 
	
	\begin{figure*}[htp!]
		\centering
		\subfloat{
			\includegraphics[width=0.45\textwidth]{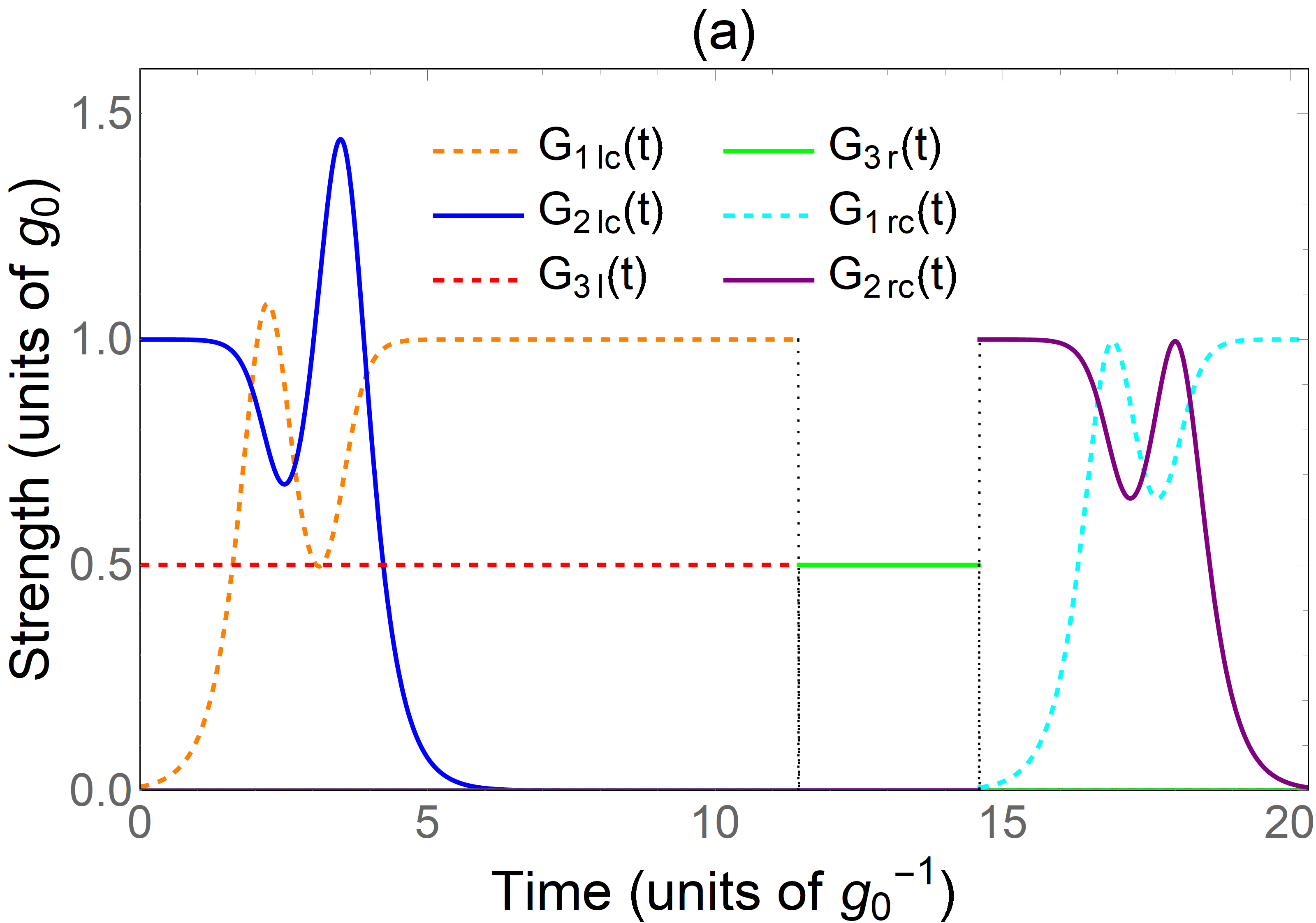}
			\label{fig2a}
		}
		\quad
		\subfloat{
			\includegraphics[width=0.45\textwidth]{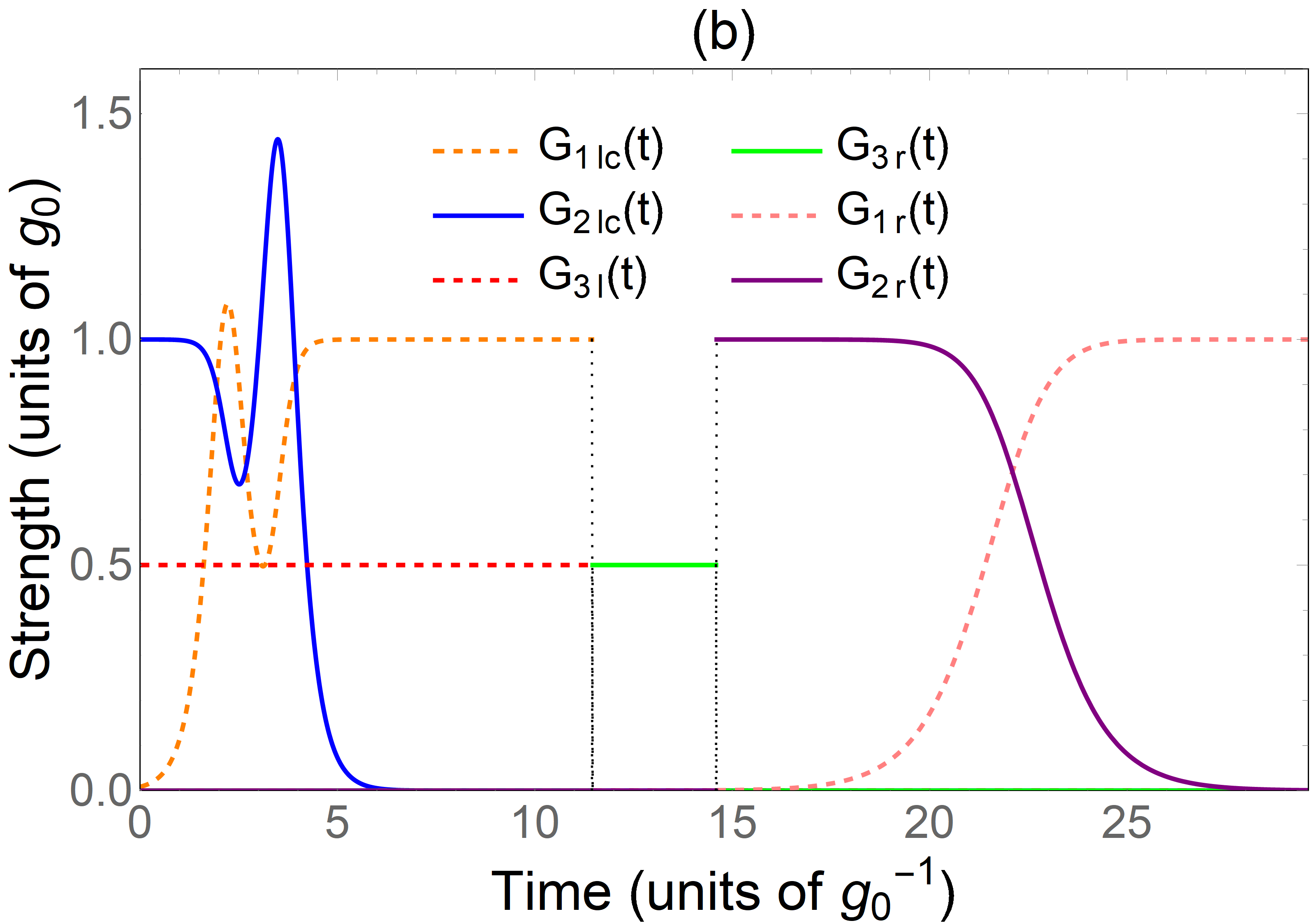}
			\label{fig2b}
		}
		\\
		\subfloat{
			\includegraphics[width=0.45\textwidth]{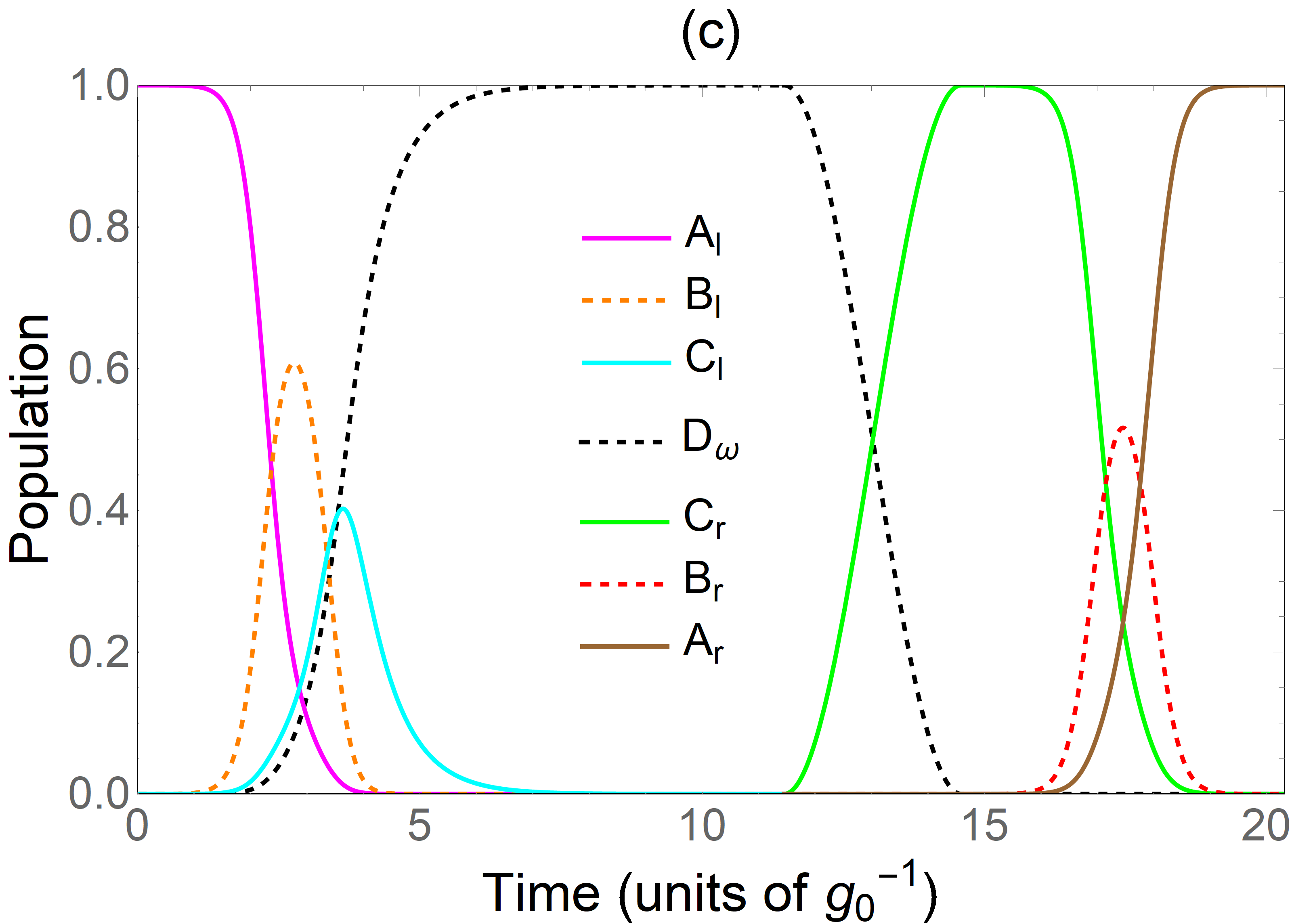}
			\label{fig2c}
		}
		\quad
		\subfloat{
			\includegraphics[width=0.45\textwidth]{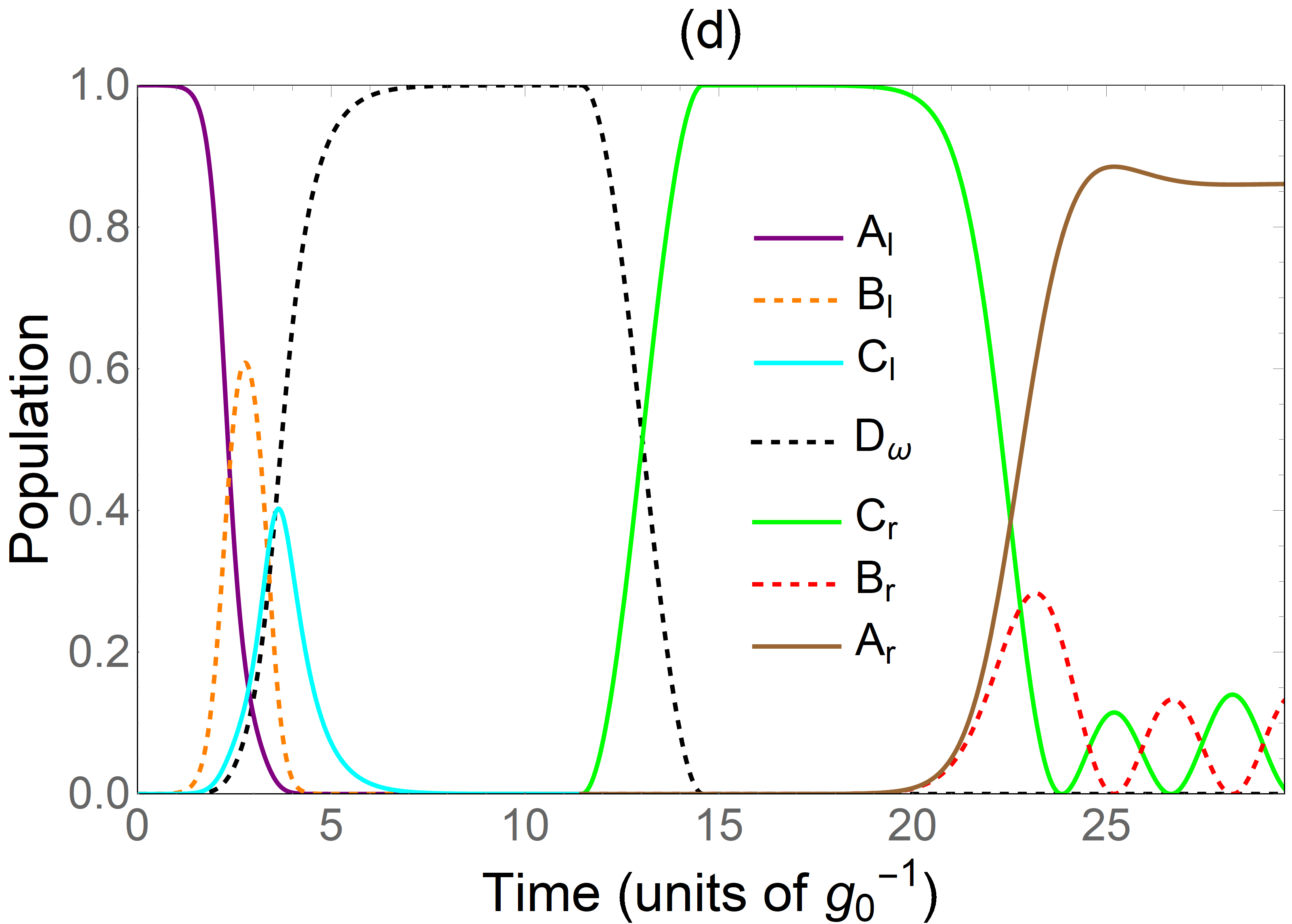}
			\label{fig2d}
		}
		\caption{\label{fig2}Comparison of the fast double-STA method and the STA-STIRAP method. The fast double-STA method contains three operations: a SATD+$\kappa$, a conversion process and a SATD. (a) Different coupling strengths in the double-STA scheme as a function of time $t$. $G_{1lc}, G_{2lc}$ (orange dashed line, blue solid line) and $G_{1rc}, G_{2rc}$ (cyan dashed line, purple solid line) are the corrected pulses strengths for SATD+$\kappa$ and SATD respectively. $G_{3l}$ (red dashed line) and $G_{3r}$ (green solid line) are the coupling strengths of different cavities and waveguide. $G_{3l}$ only turns on during the SATD+$\kappa$ and $G_{3r}$ only turns on during the conversion process with $|G_{3l}| = |G_{3r}| = 0.5 g_0$. (b) Different coupling strengths as a function of time $t$ in the STA-STIRAP scheme. (c) Population distribution of different devices depends on time $t$ in the fast doubel-STA scheme. $A_j, B_j, C_j\ (j = l, r)$ stands for the SR, NAMR and OC in node A or B, $D_\omega$ stands for the waveguide. The STA-STIRAP method consists of a SATD+$\kappa$, a conversion process and a STIRAP. (d) Population distribution of different devices depends on time $t$  in the STA-STIRAP scheme.
		}
	\end{figure*}
	
	\begin{figure}[htp!]
	    \centering
	    \includegraphics[width = 0.48\textwidth]{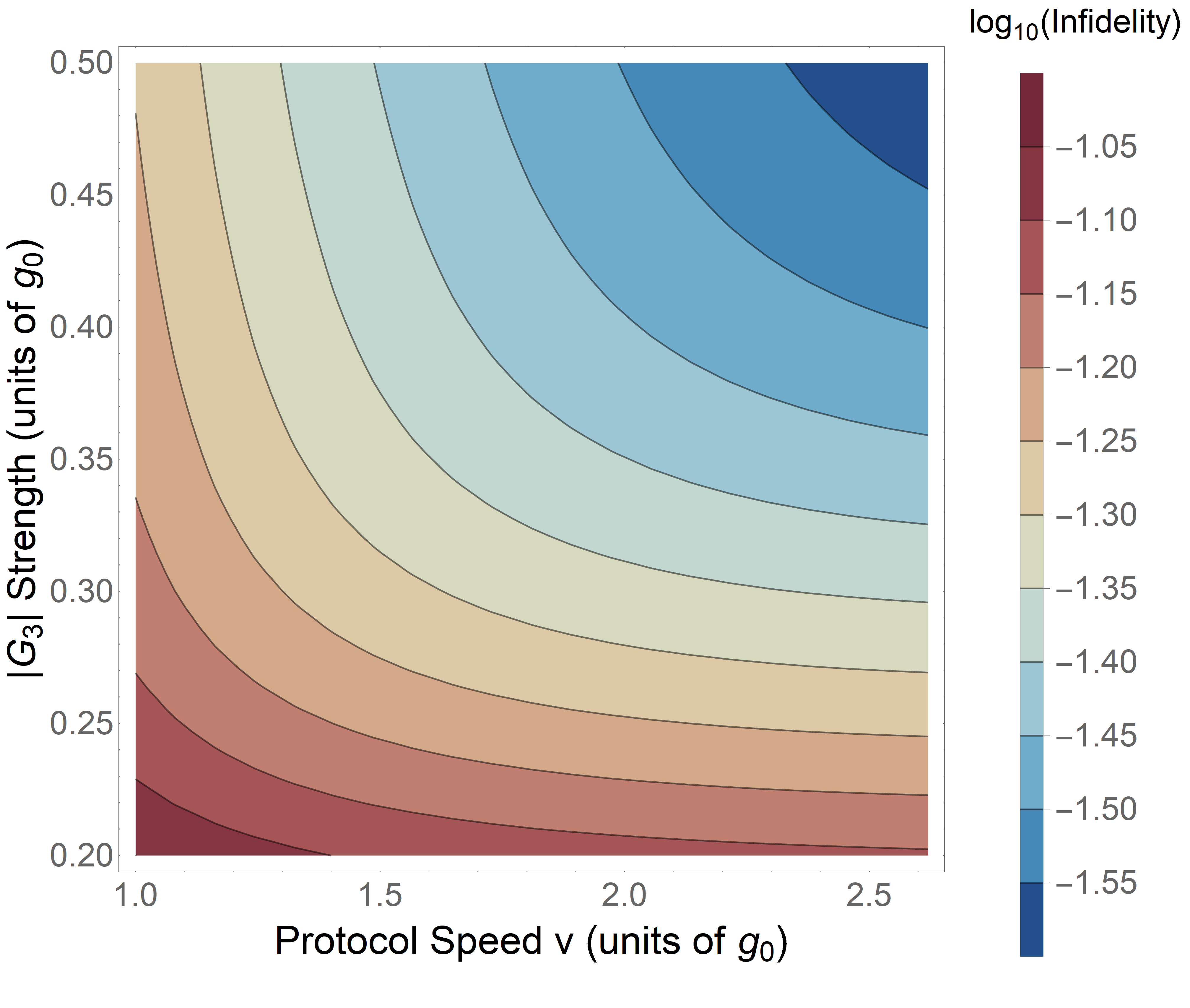}
	    \caption{Infidelity of fast double-STA scheme with dissipation depends on the protocol speed $v$ and coupling strength of cavities and waveguide $|G_3|$ (here we assume the coupling strengths of different cavities and waveguide is the same, which is $|G_3| = |G_{3l}| = |G_{3r}|$). Duration of SATD+$\kappa$ is $t_l = 15/v + 8\pi/|G_3|^2$ to make sure that state in OC completely leak to the waveguide
	    }
	    \label{fig3}
	\end{figure}
	
	\section{Numerical Simulation}
	We apply the STA part of our scheme to the optimal STIRAP pulses discussed by Vitanov \textit{et al.} in Ref. \cite{vasilev2009optimum}. The mixing angle $\theta(t)$ of the pulse is defined as follows 
	\begin{eqnarray}
		\theta(t)= \frac{\pi}{2\left(1 + e^{-vt}\right)},
	\end{eqnarray}
	where $v$ stands for the protocol speed. Here to simulate a real experiment in a lab, we truncate the pulses to a finite time interval $-t_i=t_f=7.5/v$, which ensures $G_{1lc}(t_i)=G_{1rc}(t_i')=G_{2rc}(t_f')= 10^{-3}g_0$, where $t_i$/$t_i'$ is the initial time of the pulses in node A/B and $t_f'$ is the end time of the pulses in node B. 
	
	We first use our fast double-STA approach to simulate the state transfer process without dissipation in the transfer system. Here, the protocol speeds of SATD+$\kappa$ and SATD are the same, setting as $v_l = v_r = 2.62g_0$. Fig. \ref{fig2a} shows the different coupling strengths as a function of time $t$, where the coupling strengths of OCs and the waveguide are $|G_{3l}| = |G_{3r}| = 0.5g_0$. In the SATD+$\kappa$ process, the ideally corrected \textit{Vitanov} pulses $G_{1lc}, G_{2lc}$ (orange dashed line, blue solid line), and the coupling $G_{3l}$ (red dashed line) of the OC in node A and the waveguide last for $t_l = 30/v_l$ to ensure that the population in the OC completely leaks to the waveguide. At the end of the SATD+$\kappa$ process, we turn off $G_{3l}$ and turn on the coupling $G_{3r}$ (green solid line) of the OC in node B and the waveguide, starting the conversion process. After a while $t_c = \pi/ (2|G_{3r}|)$, the states in the waveguide convert to the OC in node B, the coupling $G_{3r}$ is turned off and a SATD approach will be applied on node B. The ideally corrected \textit{Vitanov} pulses $G_{1rc}, G_{2rc}$ (cyan dashed line, purple solid line) last for the same time interval $t_f' - t_i' = 15/v_r$ due to the same protocol speed. The population distribution of the whole evolution is shown by Fig. \ref{fig2c}, where $A_j, B_j, C_j\ (j = l, r)$ stands for the SR, NAMR and OC in node A or B, and $D_\omega$ stands for the waveguide. The total fidelity of our scheme is $99.999\%$ and the time cost of these three operations is $20.32g_0^{-1}$.
	
	In order to demonstrate the advantages of our scheme, we use a STA-STIRAP scheme to transfer the state between distant interfaces. The difference between the STA-STIRAP scheme and the double-STA scheme is that we use the traditional STIRAP state transfer approach instead of the SATD approach in node B. And the STIRAP process is realized by taking the original \textit{Vitanov} pulses with the protocol speed $v_r = g_0$ . Fig. \ref{fig2b} shows the coupling strengths as a function of time $t$ and Fig. \ref{fig2d} shows the population distribution of the whole evolution depends on the time $t$. The fidelity of the STA-STIRAP state transfer scheme is $86.06\%$, and the time cost is $29.59g_0^{-1}$. Compared with the STA-STIRAP approach, our double-STA approach reaches a higher fidelity and costs less time which is more robust when the dissipation is taken into account. 
	
	Now we consider the case that the system exhibits dissipation during the state transfer process. The system Hamiltonian in this case can be easily obtained by adding non-Hermitian terms 
	\begin{eqnarray}
		H_A'(t) &=& H_A(t) - i(\gamma_{1}/2)\ket{A_l}\bra{A_r} - i(\gamma_{2}/2)\ket{B_l}\bra{B_l} \nonumber \\
		&& -i(\gamma_{3}/2)\ket{D_\omega}\bra{D_\omega}, \nonumber \\
		H_{\rm con}'(t) &=& H_{\rm con}(t) - i(\gamma_{3}/2)\ket{D_\omega}\bra{D_\omega}, \nonumber \\
		H_B'(t) &=& H_B(t) - i(\gamma_{1}/2)\ket{A_r}\bra{A_r} - i(\gamma_{2}/2)\ket{B_r}\bra{B_r}. \nonumber \\
	\end{eqnarray}
	where $\gamma_{1}$, $\gamma_{2}$ and $\gamma_{3}$ are the decay rate of SR, NAMR as well as waveguide respectively. Here, we assume that the loss of the state is caused by the waveguide when there is interaction between cavities and waveguide, so that the decay rate of cavities is neglected. Therefore, the fidelity of the SATD$+\kappa$ process in Eq. \ref{fl} becomes 
	\begin{eqnarray}
		F_l'(t)= e^{-\gamma_{3}t} \int_{t_i}^{t}d\tau \: 2\pi [G_{3l}(\tau)]^2|u_C(\tau) |^2.
	\end{eqnarray}
	
	Fig. \ref{fig3} depicts the infidelity (${\rm Infidelity} = 1 - F_{e}$) of the double-STA scheme depending on the protocol speed $v$ ($v=v_l=v_r$) and the coupling strength of the OCs and the waveguide. Here, the parameters are taken as $ \gamma_{1} = 10^{-3}g_0 $, $ \gamma_{2} = 10^{-4}g_0 $ and $ \gamma_{3} = 10^{-3}g_0 $ \cite{tittel1998experimental,teufel2011sideband,chan2011laser,pirkkalainen2013hybrid}. It can be clearly seen that reaching a higher fidelity needs both a higher protocol speed and a higher coupling strength of the OCs and the waveguide. With $v = 2.62g_0$ and $|G_3| = 0.5g_0$, the total fidelity is $97.39\%$.
	
	\section{Conclusion}
	We propose a fast high-fidelity approach to transfer state between distant optomechanical interfaces connecting by a continuum waveguide (i.e., an optical fiber). The scheme includes two parts: the local part and the remote part. In the local part, we use a SATD+$\kappa$ approach so that the state in the SR will transfer to the OC and finally leak to the waveguide. In the remote part, a population conversion process is firstly applied to convert the state from the waveguide to the OC. Then a SATD approach is used to transfer the state to the SR. Compared with the STA-STIRAP scheme, our scheme reaches a higher transfer fidelity with less time. We also apply the double-STA method to simulate the transfer evolution of the system with dissipation. In this case, a higher transfer fidelity needs higher protocol speed and higher coupling strength. Although the SATD+$\kappa$ in our protocol is irreversible, the reverse transfer process can be simply realized by inverting the pulse sequence. In a word, our scheme provides a way for state transfer on distant superconducting platforms and a way to build a quantum network.
	
	\appendix
	\bibliography{main}

\begin{thebibliography}{42}%
\makeatletter
\providecommand \@ifxundefined [1]{%
 \@ifx{#1\undefined}
}%
\providecommand \@ifnum [1]{%
 \ifnum #1\expandafter \@firstoftwo
 \else \expandafter \@secondoftwo
 \fi
}%
\providecommand \@ifx [1]{%
 \ifx #1\expandafter \@firstoftwo
 \else \expandafter \@secondoftwo
 \fi
}%
\providecommand \natexlab [1]{#1}%
\providecommand \enquote  [1]{``#1''}%
\providecommand \bibnamefont  [1]{#1}%
\providecommand \bibfnamefont [1]{#1}%
\providecommand \citenamefont [1]{#1}%
\providecommand \href@noop [0]{\@secondoftwo}%
\providecommand \href [0]{\begingroup \@sanitize@url \@href}%
\providecommand \@href[1]{\@@startlink{#1}\@@href}%
\providecommand \@@href[1]{\endgroup#1\@@endlink}%
\providecommand \@sanitize@url [0]{\catcode `\\12\catcode `\$12\catcode
  `\&12\catcode `\#12\catcode `\^12\catcode `\_12\catcode `\%12\relax}%
\providecommand \@@startlink[1]{}%
\providecommand \@@endlink[0]{}%
\providecommand \url  [0]{\begingroup\@sanitize@url \@url }%
\providecommand \@url [1]{\endgroup\@href {#1}{\urlprefix }}%
\providecommand \urlprefix  [0]{URL }%
\providecommand \Eprint [0]{\href }%
\providecommand \doibase [0]{https://doi.org/}%
\providecommand \selectlanguage [0]{\@gobble}%
\providecommand \bibinfo  [0]{\@secondoftwo}%
\providecommand \bibfield  [0]{\@secondoftwo}%
\providecommand \translation [1]{[#1]}%
\providecommand \BibitemOpen [0]{}%
\providecommand \bibitemStop [0]{}%
\providecommand \bibitemNoStop [0]{.\EOS\space}%
\providecommand \EOS [0]{\spacefactor3000\relax}%
\providecommand \BibitemShut  [1]{\csname bibitem#1\endcsname}%
\let\auto@bib@innerbib\@empty
\bibitem [{\citenamefont {Vitanov}\ \emph {et~al.}(2017)\citenamefont
  {Vitanov}, \citenamefont {Rangelov}, \citenamefont {Shore},\ and\
  \citenamefont {Bergmann}}]{vitanov2017stimulated}%
  \BibitemOpen
  \bibfield  {author} {\bibinfo {author} {\bibfnamefont {N.~V.}\ \bibnamefont
  {Vitanov}}, \bibinfo {author} {\bibfnamefont {A.~A.}\ \bibnamefont
  {Rangelov}}, \bibinfo {author} {\bibfnamefont {B.~W.}\ \bibnamefont
  {Shore}},\ and\ \bibinfo {author} {\bibfnamefont {K.}~\bibnamefont
  {Bergmann}},\ }\bibfield  {title} {\bibinfo {title} {Stimulated raman
  adiabatic passage in physics, chemistry, and beyond},\ }\href@noop {}
  {\bibfield  {journal} {\bibinfo  {journal} {Reviews of Modern Physics}\
  }\textbf {\bibinfo {volume} {89}},\ \bibinfo {pages} {015006} (\bibinfo
  {year} {2017})}\BibitemShut {NoStop}%
\bibitem [{\citenamefont {Kuhn}\ \emph {et~al.}(2019)\citenamefont {Kuhn},
  \citenamefont {Bergmann}, \citenamefont {Naegerl},\ and\ \citenamefont
  {Panda}}]{kuhn2019roadmap}%
  \BibitemOpen
  \bibfield  {author} {\bibinfo {author} {\bibfnamefont {A.}~\bibnamefont
  {Kuhn}}, \bibinfo {author} {\bibfnamefont {K.}~\bibnamefont {Bergmann}},
  \bibinfo {author} {\bibfnamefont {H.}~\bibnamefont {Naegerl}},\ and\ \bibinfo
  {author} {\bibfnamefont {C.}~\bibnamefont {Panda}},\ }\bibfield  {title}
  {\bibinfo {title} {Roadmap on stirap applications},\ }\href@noop {}
  {\bibfield  {journal} {\bibinfo  {journal} {Journal of Physics B: Atomic,
  Molecular and Optical Physics}\ }\textbf {\bibinfo {volume} {52}} (\bibinfo
  {year} {2019})}\BibitemShut {NoStop}%
\bibitem [{\citenamefont {Kulin}\ \emph {et~al.}(1997)\citenamefont {Kulin},
  \citenamefont {Saubamea}, \citenamefont {Peik}, \citenamefont {Lawall},
  \citenamefont {Hijmans}, \citenamefont {Leduc},\ and\ \citenamefont
  {Cohen-Tannoudji}}]{kulin1997coherent}%
  \BibitemOpen
  \bibfield  {author} {\bibinfo {author} {\bibfnamefont {S.}~\bibnamefont
  {Kulin}}, \bibinfo {author} {\bibfnamefont {B.}~\bibnamefont {Saubamea}},
  \bibinfo {author} {\bibfnamefont {E.}~\bibnamefont {Peik}}, \bibinfo {author}
  {\bibfnamefont {J.}~\bibnamefont {Lawall}}, \bibinfo {author} {\bibfnamefont
  {T.}~\bibnamefont {Hijmans}}, \bibinfo {author} {\bibfnamefont
  {M.}~\bibnamefont {Leduc}},\ and\ \bibinfo {author} {\bibfnamefont
  {C.}~\bibnamefont {Cohen-Tannoudji}},\ }\bibfield  {title} {\bibinfo {title}
  {Coherent manipulation of atomic wave packets by adiabatic transfer},\
  }\href@noop {} {\bibfield  {journal} {\bibinfo  {journal} {Physical review
  letters}\ }\textbf {\bibinfo {volume} {78}},\ \bibinfo {pages} {4185}
  (\bibinfo {year} {1997})}\BibitemShut {NoStop}%
\bibitem [{\citenamefont {Theuer}\ and\ \citenamefont
  {Bergmann}(1998)}]{theuer1998atomic}%
  \BibitemOpen
  \bibfield  {author} {\bibinfo {author} {\bibfnamefont {H.}~\bibnamefont
  {Theuer}}\ and\ \bibinfo {author} {\bibfnamefont {K.}~\bibnamefont
  {Bergmann}},\ }\bibfield  {title} {\bibinfo {title} {Atomic beam deflection
  by coherent momentum transfer and the dependence on weak magnetic fields},\
  }\href@noop {} {\bibfield  {journal} {\bibinfo  {journal} {The European
  Physical Journal D-Atomic, Molecular, Optical and Plasma Physics}\ }\textbf
  {\bibinfo {volume} {2}},\ \bibinfo {pages} {279} (\bibinfo {year}
  {1998})}\BibitemShut {NoStop}%
\bibitem [{\citenamefont {N{\"o}lleke}\ \emph {et~al.}(2013)\citenamefont
  {N{\"o}lleke}, \citenamefont {Neuzner}, \citenamefont {Reiserer},
  \citenamefont {Hahn}, \citenamefont {Rempe},\ and\ \citenamefont
  {Ritter}}]{nolleke2013efficient}%
  \BibitemOpen
  \bibfield  {author} {\bibinfo {author} {\bibfnamefont {C.}~\bibnamefont
  {N{\"o}lleke}}, \bibinfo {author} {\bibfnamefont {A.}~\bibnamefont
  {Neuzner}}, \bibinfo {author} {\bibfnamefont {A.}~\bibnamefont {Reiserer}},
  \bibinfo {author} {\bibfnamefont {C.}~\bibnamefont {Hahn}}, \bibinfo {author}
  {\bibfnamefont {G.}~\bibnamefont {Rempe}},\ and\ \bibinfo {author}
  {\bibfnamefont {S.}~\bibnamefont {Ritter}},\ }\bibfield  {title} {\bibinfo
  {title} {Efficient teleportation between remote single-atom quantum
  memories},\ }\href@noop {} {\bibfield  {journal} {\bibinfo  {journal}
  {Physical review letters}\ }\textbf {\bibinfo {volume} {110}},\ \bibinfo
  {pages} {140403} (\bibinfo {year} {2013})}\BibitemShut {NoStop}%
\bibitem [{\citenamefont {Takekoshi}\ \emph {et~al.}(2014)\citenamefont
  {Takekoshi}, \citenamefont {Reichs{\"o}llner}, \citenamefont {Schindewolf},
  \citenamefont {Hutson}, \citenamefont {Le~Sueur}, \citenamefont {Dulieu},
  \citenamefont {Ferlaino}, \citenamefont {Grimm},\ and\ \citenamefont
  {N{\"a}gerl}}]{takekoshi2014ultracold}%
  \BibitemOpen
  \bibfield  {author} {\bibinfo {author} {\bibfnamefont {T.}~\bibnamefont
  {Takekoshi}}, \bibinfo {author} {\bibfnamefont {L.}~\bibnamefont
  {Reichs{\"o}llner}}, \bibinfo {author} {\bibfnamefont {A.}~\bibnamefont
  {Schindewolf}}, \bibinfo {author} {\bibfnamefont {J.~M.}\ \bibnamefont
  {Hutson}}, \bibinfo {author} {\bibfnamefont {C.~R.}\ \bibnamefont
  {Le~Sueur}}, \bibinfo {author} {\bibfnamefont {O.}~\bibnamefont {Dulieu}},
  \bibinfo {author} {\bibfnamefont {F.}~\bibnamefont {Ferlaino}}, \bibinfo
  {author} {\bibfnamefont {R.}~\bibnamefont {Grimm}},\ and\ \bibinfo {author}
  {\bibfnamefont {H.-C.}\ \bibnamefont {N{\"a}gerl}},\ }\bibfield  {title}
  {\bibinfo {title} {Ultracold dense samples of dipolar rbcs molecules in the
  rovibrational and hyperfine ground state},\ }\href@noop {} {\bibfield
  {journal} {\bibinfo  {journal} {Physical review letters}\ }\textbf {\bibinfo
  {volume} {113}},\ \bibinfo {pages} {205301} (\bibinfo {year}
  {2014})}\BibitemShut {NoStop}%
\bibitem [{\citenamefont {Beterov}\ \emph {et~al.}(2013)\citenamefont
  {Beterov}, \citenamefont {Saffman}, \citenamefont {Yakshina}, \citenamefont
  {Zhukov}, \citenamefont {Tretyakov}, \citenamefont {Entin}, \citenamefont
  {Ryabtsev}, \citenamefont {Mansell}, \citenamefont {MacCormick},
  \citenamefont {Bergamini} \emph {et~al.}}]{beterov2013quantum}%
  \BibitemOpen
  \bibfield  {author} {\bibinfo {author} {\bibfnamefont {I.}~\bibnamefont
  {Beterov}}, \bibinfo {author} {\bibfnamefont {M.}~\bibnamefont {Saffman}},
  \bibinfo {author} {\bibfnamefont {E.}~\bibnamefont {Yakshina}}, \bibinfo
  {author} {\bibfnamefont {V.}~\bibnamefont {Zhukov}}, \bibinfo {author}
  {\bibfnamefont {D.}~\bibnamefont {Tretyakov}}, \bibinfo {author}
  {\bibfnamefont {V.}~\bibnamefont {Entin}}, \bibinfo {author} {\bibfnamefont
  {I.}~\bibnamefont {Ryabtsev}}, \bibinfo {author} {\bibfnamefont
  {C.}~\bibnamefont {Mansell}}, \bibinfo {author} {\bibfnamefont
  {C.}~\bibnamefont {MacCormick}}, \bibinfo {author} {\bibfnamefont
  {S.}~\bibnamefont {Bergamini}}, \emph {et~al.},\ }\bibfield  {title}
  {\bibinfo {title} {Quantum gates in mesoscopic atomic ensembles based on
  adiabatic passage and rydberg blockade},\ }\href@noop {} {\bibfield
  {journal} {\bibinfo  {journal} {Physical Review A}\ }\textbf {\bibinfo
  {volume} {88}},\ \bibinfo {pages} {010303} (\bibinfo {year}
  {2013})}\BibitemShut {NoStop}%
\bibitem [{\citenamefont {Noguchi}\ \emph {et~al.}(2012)\citenamefont
  {Noguchi}, \citenamefont {Toyoda},\ and\ \citenamefont
  {Urabe}}]{noguchi2012generation}%
  \BibitemOpen
  \bibfield  {author} {\bibinfo {author} {\bibfnamefont {A.}~\bibnamefont
  {Noguchi}}, \bibinfo {author} {\bibfnamefont {K.}~\bibnamefont {Toyoda}},\
  and\ \bibinfo {author} {\bibfnamefont {S.}~\bibnamefont {Urabe}},\ }\bibfield
   {title} {\bibinfo {title} {Generation of dicke states with phonon-mediated
  multilevel stimulated raman adiabatic passage},\ }\href@noop {} {\bibfield
  {journal} {\bibinfo  {journal} {Physical review letters}\ }\textbf {\bibinfo
  {volume} {109}},\ \bibinfo {pages} {260502} (\bibinfo {year}
  {2012})}\BibitemShut {NoStop}%
\bibitem [{\citenamefont {Golter}\ \emph {et~al.}(2016)\citenamefont {Golter},
  \citenamefont {Oo}, \citenamefont {Amezcua}, \citenamefont {Stewart},\ and\
  \citenamefont {Wang}}]{golter2016optomechanical}%
  \BibitemOpen
  \bibfield  {author} {\bibinfo {author} {\bibfnamefont {D.~A.}\ \bibnamefont
  {Golter}}, \bibinfo {author} {\bibfnamefont {T.}~\bibnamefont {Oo}}, \bibinfo
  {author} {\bibfnamefont {M.}~\bibnamefont {Amezcua}}, \bibinfo {author}
  {\bibfnamefont {K.~A.}\ \bibnamefont {Stewart}},\ and\ \bibinfo {author}
  {\bibfnamefont {H.}~\bibnamefont {Wang}},\ }\bibfield  {title} {\bibinfo
  {title} {Optomechanical quantum control of a nitrogen-vacancy center in
  diamond},\ }\href@noop {} {\bibfield  {journal} {\bibinfo  {journal}
  {Physical review letters}\ }\textbf {\bibinfo {volume} {116}},\ \bibinfo
  {pages} {143602} (\bibinfo {year} {2016})}\BibitemShut {NoStop}%
\bibitem [{\citenamefont {Kumar}\ \emph {et~al.}(2016)\citenamefont {Kumar},
  \citenamefont {Veps{\"a}l{\"a}inen}, \citenamefont {Danilin},\ and\
  \citenamefont {Paraoanu}}]{kumar2016stimulated}%
  \BibitemOpen
  \bibfield  {author} {\bibinfo {author} {\bibfnamefont {K.}~\bibnamefont
  {Kumar}}, \bibinfo {author} {\bibfnamefont {A.}~\bibnamefont
  {Veps{\"a}l{\"a}inen}}, \bibinfo {author} {\bibfnamefont {S.}~\bibnamefont
  {Danilin}},\ and\ \bibinfo {author} {\bibfnamefont {G.}~\bibnamefont
  {Paraoanu}},\ }\bibfield  {title} {\bibinfo {title} {Stimulated raman
  adiabatic passage in a three-level superconducting circuit},\ }\href@noop {}
  {\bibfield  {journal} {\bibinfo  {journal} {Nature communications}\ }\textbf
  {\bibinfo {volume} {7}},\ \bibinfo {pages} {1} (\bibinfo {year}
  {2016})}\BibitemShut {NoStop}%
\bibitem [{\citenamefont {Xu}\ \emph {et~al.}(2016)\citenamefont {Xu},
  \citenamefont {Song}, \citenamefont {Liu}, \citenamefont {Xue}, \citenamefont
  {Su}, \citenamefont {Deng}, \citenamefont {Tian}, \citenamefont {Zheng},
  \citenamefont {Han}, \citenamefont {Zhong} \emph {et~al.}}]{xu2016coherent}%
  \BibitemOpen
  \bibfield  {author} {\bibinfo {author} {\bibfnamefont {H.}~\bibnamefont
  {Xu}}, \bibinfo {author} {\bibfnamefont {C.}~\bibnamefont {Song}}, \bibinfo
  {author} {\bibfnamefont {W.}~\bibnamefont {Liu}}, \bibinfo {author}
  {\bibfnamefont {G.}~\bibnamefont {Xue}}, \bibinfo {author} {\bibfnamefont
  {F.}~\bibnamefont {Su}}, \bibinfo {author} {\bibfnamefont {H.}~\bibnamefont
  {Deng}}, \bibinfo {author} {\bibfnamefont {Y.}~\bibnamefont {Tian}}, \bibinfo
  {author} {\bibfnamefont {D.}~\bibnamefont {Zheng}}, \bibinfo {author}
  {\bibfnamefont {S.}~\bibnamefont {Han}}, \bibinfo {author} {\bibfnamefont
  {Y.-P.}\ \bibnamefont {Zhong}}, \emph {et~al.},\ }\bibfield  {title}
  {\bibinfo {title} {Coherent population transfer between uncoupled or weakly
  coupled states in ladder-type superconducting qutrits},\ }\href@noop {}
  {\bibfield  {journal} {\bibinfo  {journal} {Nature communications}\ }\textbf
  {\bibinfo {volume} {7}},\ \bibinfo {pages} {1} (\bibinfo {year}
  {2016})}\BibitemShut {NoStop}%
\bibitem [{\citenamefont {Simon}\ \emph {et~al.}(2011)\citenamefont {Simon},
  \citenamefont {Belhadj}, \citenamefont {Chatel}, \citenamefont {Amand},
  \citenamefont {Renucci}, \citenamefont {Lema{\^\i}tre}, \citenamefont
  {Krebs}, \citenamefont {Dalgarno}, \citenamefont {Warburton}, \citenamefont
  {Marie} \emph {et~al.}}]{simon2011robust}%
  \BibitemOpen
  \bibfield  {author} {\bibinfo {author} {\bibfnamefont {C.-M.}\ \bibnamefont
  {Simon}}, \bibinfo {author} {\bibfnamefont {T.}~\bibnamefont {Belhadj}},
  \bibinfo {author} {\bibfnamefont {B.}~\bibnamefont {Chatel}}, \bibinfo
  {author} {\bibfnamefont {T.}~\bibnamefont {Amand}}, \bibinfo {author}
  {\bibfnamefont {P.}~\bibnamefont {Renucci}}, \bibinfo {author} {\bibfnamefont
  {A.}~\bibnamefont {Lema{\^\i}tre}}, \bibinfo {author} {\bibfnamefont
  {O.}~\bibnamefont {Krebs}}, \bibinfo {author} {\bibfnamefont
  {P.}~\bibnamefont {Dalgarno}}, \bibinfo {author} {\bibfnamefont
  {R.}~\bibnamefont {Warburton}}, \bibinfo {author} {\bibfnamefont
  {X.}~\bibnamefont {Marie}}, \emph {et~al.},\ }\bibfield  {title} {\bibinfo
  {title} {Robust quantum dot exciton generation via adiabatic passage with
  frequency-swept optical pulses},\ }\href@noop {} {\bibfield  {journal}
  {\bibinfo  {journal} {Physical review letters}\ }\textbf {\bibinfo {volume}
  {106}},\ \bibinfo {pages} {166801} (\bibinfo {year} {2011})}\BibitemShut
  {NoStop}%
\bibitem [{\citenamefont {Tomaino}\ \emph {et~al.}(2012)\citenamefont
  {Tomaino}, \citenamefont {Jameson}, \citenamefont {Lee}, \citenamefont
  {Khitrova}, \citenamefont {Gibbs}, \citenamefont {Klettke}, \citenamefont
  {Kira},\ and\ \citenamefont {Koch}}]{tomaino2012terahertz}%
  \BibitemOpen
  \bibfield  {author} {\bibinfo {author} {\bibfnamefont {J.}~\bibnamefont
  {Tomaino}}, \bibinfo {author} {\bibfnamefont {A.}~\bibnamefont {Jameson}},
  \bibinfo {author} {\bibfnamefont {Y.-S.}\ \bibnamefont {Lee}}, \bibinfo
  {author} {\bibfnamefont {G.}~\bibnamefont {Khitrova}}, \bibinfo {author}
  {\bibfnamefont {H.}~\bibnamefont {Gibbs}}, \bibinfo {author} {\bibfnamefont
  {A.}~\bibnamefont {Klettke}}, \bibinfo {author} {\bibfnamefont
  {M.}~\bibnamefont {Kira}},\ and\ \bibinfo {author} {\bibfnamefont
  {S.}~\bibnamefont {Koch}},\ }\bibfield  {title} {\bibinfo {title} {Terahertz
  excitation of a coherent $\lambda$-type three-level system of
  exciton-polariton modes in a quantum-well microcavity},\ }\href@noop {}
  {\bibfield  {journal} {\bibinfo  {journal} {Physical review letters}\
  }\textbf {\bibinfo {volume} {108}},\ \bibinfo {pages} {267402} (\bibinfo
  {year} {2012})}\BibitemShut {NoStop}%
\bibitem [{\citenamefont {Gu{\'e}ry-Odelin}\ \emph {et~al.}(2019)\citenamefont
  {Gu{\'e}ry-Odelin}, \citenamefont {Ruschhaupt}, \citenamefont {Kiely},
  \citenamefont {Torrontegui}, \citenamefont {Mart{\'\i}nez-Garaot},\ and\
  \citenamefont {Muga}}]{guery2019shortcuts}%
  \BibitemOpen
  \bibfield  {author} {\bibinfo {author} {\bibfnamefont {D.}~\bibnamefont
  {Gu{\'e}ry-Odelin}}, \bibinfo {author} {\bibfnamefont {A.}~\bibnamefont
  {Ruschhaupt}}, \bibinfo {author} {\bibfnamefont {A.}~\bibnamefont {Kiely}},
  \bibinfo {author} {\bibfnamefont {E.}~\bibnamefont {Torrontegui}}, \bibinfo
  {author} {\bibfnamefont {S.}~\bibnamefont {Mart{\'\i}nez-Garaot}},\ and\
  \bibinfo {author} {\bibfnamefont {J.~G.}\ \bibnamefont {Muga}},\ }\bibfield
  {title} {\bibinfo {title} {Shortcuts to adiabaticity: Concepts, methods, and
  applications},\ }\href@noop {} {\bibfield  {journal} {\bibinfo  {journal}
  {Reviews of Modern Physics}\ }\textbf {\bibinfo {volume} {91}},\ \bibinfo
  {pages} {045001} (\bibinfo {year} {2019})}\BibitemShut {NoStop}%
\bibitem [{\citenamefont {Baksic}\ \emph {et~al.}(2016)\citenamefont {Baksic},
  \citenamefont {Ribeiro},\ and\ \citenamefont {Clerk}}]{baksic2016speeding}%
  \BibitemOpen
  \bibfield  {author} {\bibinfo {author} {\bibfnamefont {A.}~\bibnamefont
  {Baksic}}, \bibinfo {author} {\bibfnamefont {H.}~\bibnamefont {Ribeiro}},\
  and\ \bibinfo {author} {\bibfnamefont {A.~A.}\ \bibnamefont {Clerk}},\
  }\bibfield  {title} {\bibinfo {title} {Speeding up adiabatic quantum state
  transfer by using dressed states},\ }\href@noop {} {\bibfield  {journal}
  {\bibinfo  {journal} {Physical review letters}\ }\textbf {\bibinfo {volume}
  {116}},\ \bibinfo {pages} {230503} (\bibinfo {year} {2016})}\BibitemShut
  {NoStop}%
\bibitem [{\citenamefont {Sels}\ and\ \citenamefont
  {Polkovnikov}(2017)}]{sels2017minimizing}%
  \BibitemOpen
  \bibfield  {author} {\bibinfo {author} {\bibfnamefont {D.}~\bibnamefont
  {Sels}}\ and\ \bibinfo {author} {\bibfnamefont {A.}~\bibnamefont
  {Polkovnikov}},\ }\bibfield  {title} {\bibinfo {title} {Minimizing
  irreversible losses in quantum systems by local counterdiabatic driving},\
  }\href@noop {} {\bibfield  {journal} {\bibinfo  {journal} {Proceedings of the
  National Academy of Sciences}\ }\textbf {\bibinfo {volume} {114}},\ \bibinfo
  {pages} {E3909} (\bibinfo {year} {2017})}\BibitemShut {NoStop}%
\bibitem [{\citenamefont {Petiziol}\ \emph {et~al.}(2018)\citenamefont
  {Petiziol}, \citenamefont {Dive}, \citenamefont {Mintert},\ and\
  \citenamefont {Wimberger}}]{petiziol2018fast}%
  \BibitemOpen
  \bibfield  {author} {\bibinfo {author} {\bibfnamefont {F.}~\bibnamefont
  {Petiziol}}, \bibinfo {author} {\bibfnamefont {B.}~\bibnamefont {Dive}},
  \bibinfo {author} {\bibfnamefont {F.}~\bibnamefont {Mintert}},\ and\ \bibinfo
  {author} {\bibfnamefont {S.}~\bibnamefont {Wimberger}},\ }\bibfield  {title}
  {\bibinfo {title} {Fast adiabatic evolution by oscillating initial
  hamiltonians},\ }\href@noop {} {\bibfield  {journal} {\bibinfo  {journal}
  {Physical Review A}\ }\textbf {\bibinfo {volume} {98}},\ \bibinfo {pages}
  {043436} (\bibinfo {year} {2018})}\BibitemShut {NoStop}%
\bibitem [{\citenamefont {Kiely}\ and\ \citenamefont
  {Ruschhaupt}(2014)}]{kiely2014inhibiting}%
  \BibitemOpen
  \bibfield  {author} {\bibinfo {author} {\bibfnamefont {A.}~\bibnamefont
  {Kiely}}\ and\ \bibinfo {author} {\bibfnamefont {A.}~\bibnamefont
  {Ruschhaupt}},\ }\bibfield  {title} {\bibinfo {title} {Inhibiting unwanted
  transitions in population transfer in two-and three-level quantum systems},\
  }\href@noop {} {\bibfield  {journal} {\bibinfo  {journal} {Journal of Physics
  B: Atomic, Molecular and Optical Physics}\ }\textbf {\bibinfo {volume}
  {47}},\ \bibinfo {pages} {115501} (\bibinfo {year} {2014})}\BibitemShut
  {NoStop}%
\bibitem [{\citenamefont {Kiran}\ and\ \citenamefont
  {Ponmurugan}(2021)}]{kiran2021invariant}%
  \BibitemOpen
  \bibfield  {author} {\bibinfo {author} {\bibfnamefont {T.}~\bibnamefont
  {Kiran}}\ and\ \bibinfo {author} {\bibfnamefont {M.}~\bibnamefont
  {Ponmurugan}},\ }\bibfield  {title} {\bibinfo {title} {Invariant-based
  investigation of shortcut to adiabaticity for quantum harmonic oscillators
  under a time-varying frictional force},\ }\href@noop {} {\bibfield  {journal}
  {\bibinfo  {journal} {Physical Review A}\ }\textbf {\bibinfo {volume}
  {103}},\ \bibinfo {pages} {042206} (\bibinfo {year} {2021})}\BibitemShut
  {NoStop}%
\bibitem [{\citenamefont {Han}\ \emph {et~al.}(2021)\citenamefont {Han},
  \citenamefont {Dong}, \citenamefont {Yang}, \citenamefont {Song},
  \citenamefont {Qiu}, \citenamefont {Zheng}, \citenamefont {Xu}, \citenamefont
  {Huang}, \citenamefont {Wang}, \citenamefont {Lan} \emph
  {et~al.}}]{han2021realization}%
  \BibitemOpen
  \bibfield  {author} {\bibinfo {author} {\bibfnamefont {Z.}~\bibnamefont
  {Han}}, \bibinfo {author} {\bibfnamefont {Y.}~\bibnamefont {Dong}}, \bibinfo
  {author} {\bibfnamefont {X.}~\bibnamefont {Yang}}, \bibinfo {author}
  {\bibfnamefont {S.}~\bibnamefont {Song}}, \bibinfo {author} {\bibfnamefont
  {L.}~\bibnamefont {Qiu}}, \bibinfo {author} {\bibfnamefont {W.}~\bibnamefont
  {Zheng}}, \bibinfo {author} {\bibfnamefont {J.}~\bibnamefont {Xu}}, \bibinfo
  {author} {\bibfnamefont {T.}~\bibnamefont {Huang}}, \bibinfo {author}
  {\bibfnamefont {Z.}~\bibnamefont {Wang}}, \bibinfo {author} {\bibfnamefont
  {D.}~\bibnamefont {Lan}}, \emph {et~al.},\ }\bibfield  {title} {\bibinfo
  {title} {Realization of invariant-based shortcuts to population inversion
  with a superconducting circuit},\ }\href@noop {} {\bibfield  {journal}
  {\bibinfo  {journal} {Applied Physics Letters}\ }\textbf {\bibinfo {volume}
  {118}},\ \bibinfo {pages} {224003} (\bibinfo {year} {2021})}\BibitemShut
  {NoStop}%
\bibitem [{\citenamefont {Masuda}\ and\ \citenamefont
  {Nakamura}(2010)}]{masuda2010fast}%
  \BibitemOpen
  \bibfield  {author} {\bibinfo {author} {\bibfnamefont {S.}~\bibnamefont
  {Masuda}}\ and\ \bibinfo {author} {\bibfnamefont {K.}~\bibnamefont
  {Nakamura}},\ }\bibfield  {title} {\bibinfo {title} {Fast-forward of
  adiabatic dynamics in quantum mechanics},\ }\href@noop {} {\bibfield
  {journal} {\bibinfo  {journal} {Proceedings of the Royal Society A:
  Mathematical, Physical and Engineering Sciences}\ }\textbf {\bibinfo {volume}
  {466}},\ \bibinfo {pages} {1135} (\bibinfo {year} {2010})}\BibitemShut
  {NoStop}%
\bibitem [{\citenamefont {Torrontegui}\ \emph {et~al.}(2012)\citenamefont
  {Torrontegui}, \citenamefont {Mart{\'\i}nez-Garaot}, \citenamefont
  {Ruschhaupt},\ and\ \citenamefont {Muga}}]{torrontegui2012shortcuts}%
  \BibitemOpen
  \bibfield  {author} {\bibinfo {author} {\bibfnamefont {E.}~\bibnamefont
  {Torrontegui}}, \bibinfo {author} {\bibfnamefont {S.}~\bibnamefont
  {Mart{\'\i}nez-Garaot}}, \bibinfo {author} {\bibfnamefont {A.}~\bibnamefont
  {Ruschhaupt}},\ and\ \bibinfo {author} {\bibfnamefont {J.~G.}\ \bibnamefont
  {Muga}},\ }\bibfield  {title} {\bibinfo {title} {Shortcuts to adiabaticity:
  fast-forward approach},\ }\href@noop {} {\bibfield  {journal} {\bibinfo
  {journal} {Physical Review A}\ }\textbf {\bibinfo {volume} {86}},\ \bibinfo
  {pages} {013601} (\bibinfo {year} {2012})}\BibitemShut {NoStop}%
\bibitem [{\citenamefont {Patra}\ and\ \citenamefont
  {Jarzynski}(2021)}]{patra2021semiclassical}%
  \BibitemOpen
  \bibfield  {author} {\bibinfo {author} {\bibfnamefont {A.}~\bibnamefont
  {Patra}}\ and\ \bibinfo {author} {\bibfnamefont {C.}~\bibnamefont
  {Jarzynski}},\ }\bibfield  {title} {\bibinfo {title} {Semiclassical
  fast-forward shortcuts to adiabaticity},\ }\href@noop {} {\bibfield
  {journal} {\bibinfo  {journal} {Physical Review Research}\ }\textbf {\bibinfo
  {volume} {3}},\ \bibinfo {pages} {013087} (\bibinfo {year}
  {2021})}\BibitemShut {NoStop}%
\bibitem [{\citenamefont {Faure}\ \emph {et~al.}(2019)\citenamefont {Faure},
  \citenamefont {Ciliberto}, \citenamefont {Trizac},\ and\ \citenamefont
  {Gu{\'e}ry-Odelin}}]{faure2019shortcut}%
  \BibitemOpen
  \bibfield  {author} {\bibinfo {author} {\bibfnamefont {S.}~\bibnamefont
  {Faure}}, \bibinfo {author} {\bibfnamefont {S.}~\bibnamefont {Ciliberto}},
  \bibinfo {author} {\bibfnamefont {E.}~\bibnamefont {Trizac}},\ and\ \bibinfo
  {author} {\bibfnamefont {D.}~\bibnamefont {Gu{\'e}ry-Odelin}},\ }\bibfield
  {title} {\bibinfo {title} {Shortcut to stationary regimes: A simple
  experimental demonstration},\ }\href@noop {} {\bibfield  {journal} {\bibinfo
  {journal} {American Journal of Physics}\ }\textbf {\bibinfo {volume} {87}},\
  \bibinfo {pages} {125} (\bibinfo {year} {2019})}\BibitemShut {NoStop}%
\bibitem [{\citenamefont {Veps{\"a}l{\"a}inen}\ \emph
  {et~al.}(2019)\citenamefont {Veps{\"a}l{\"a}inen}, \citenamefont {Danilin},\
  and\ \citenamefont {Paraoanu}}]{vepsalainen2019superadiabatic}%
  \BibitemOpen
  \bibfield  {author} {\bibinfo {author} {\bibfnamefont {A.}~\bibnamefont
  {Veps{\"a}l{\"a}inen}}, \bibinfo {author} {\bibfnamefont {S.}~\bibnamefont
  {Danilin}},\ and\ \bibinfo {author} {\bibfnamefont {G.~S.}\ \bibnamefont
  {Paraoanu}},\ }\bibfield  {title} {\bibinfo {title} {Superadiabatic
  population transfer in a three-level superconducting circuit},\ }\href@noop
  {} {\bibfield  {journal} {\bibinfo  {journal} {Science advances}\ }\textbf
  {\bibinfo {volume} {5}},\ \bibinfo {pages} {eaau5999} (\bibinfo {year}
  {2019})}\BibitemShut {NoStop}%
\bibitem [{\citenamefont {Yan}\ \emph {et~al.}(2019)\citenamefont {Yan},
  \citenamefont {Liu}, \citenamefont {Xu}, \citenamefont {Song}, \citenamefont
  {Liu}, \citenamefont {Zhang}, \citenamefont {Deng}, \citenamefont {Yan},
  \citenamefont {Rong}, \citenamefont {Huang} \emph
  {et~al.}}]{yan2019experimental}%
  \BibitemOpen
  \bibfield  {author} {\bibinfo {author} {\bibfnamefont {T.}~\bibnamefont
  {Yan}}, \bibinfo {author} {\bibfnamefont {B.-J.}\ \bibnamefont {Liu}},
  \bibinfo {author} {\bibfnamefont {K.}~\bibnamefont {Xu}}, \bibinfo {author}
  {\bibfnamefont {C.}~\bibnamefont {Song}}, \bibinfo {author} {\bibfnamefont
  {S.}~\bibnamefont {Liu}}, \bibinfo {author} {\bibfnamefont {Z.}~\bibnamefont
  {Zhang}}, \bibinfo {author} {\bibfnamefont {H.}~\bibnamefont {Deng}},
  \bibinfo {author} {\bibfnamefont {Z.}~\bibnamefont {Yan}}, \bibinfo {author}
  {\bibfnamefont {H.}~\bibnamefont {Rong}}, \bibinfo {author} {\bibfnamefont
  {K.}~\bibnamefont {Huang}}, \emph {et~al.},\ }\bibfield  {title} {\bibinfo
  {title} {Experimental realization of nonadiabatic shortcut to non-abelian
  geometric gates},\ }\href@noop {} {\bibfield  {journal} {\bibinfo  {journal}
  {Physical review letters}\ }\textbf {\bibinfo {volume} {122}},\ \bibinfo
  {pages} {080501} (\bibinfo {year} {2019})}\BibitemShut {NoStop}%
\bibitem [{\citenamefont {Qiu}\ \emph {et~al.}(2021)\citenamefont {Qiu},
  \citenamefont {Li}, \citenamefont {Han}, \citenamefont {Zheng}, \citenamefont
  {Yang}, \citenamefont {Dong}, \citenamefont {Song}, \citenamefont {Lan},
  \citenamefont {Tan},\ and\ \citenamefont {Yu}}]{qiu2021experimental}%
  \BibitemOpen
  \bibfield  {author} {\bibinfo {author} {\bibfnamefont {L.}~\bibnamefont
  {Qiu}}, \bibinfo {author} {\bibfnamefont {H.}~\bibnamefont {Li}}, \bibinfo
  {author} {\bibfnamefont {Z.}~\bibnamefont {Han}}, \bibinfo {author}
  {\bibfnamefont {W.}~\bibnamefont {Zheng}}, \bibinfo {author} {\bibfnamefont
  {X.}~\bibnamefont {Yang}}, \bibinfo {author} {\bibfnamefont {Y.}~\bibnamefont
  {Dong}}, \bibinfo {author} {\bibfnamefont {S.}~\bibnamefont {Song}}, \bibinfo
  {author} {\bibfnamefont {D.}~\bibnamefont {Lan}}, \bibinfo {author}
  {\bibfnamefont {X.}~\bibnamefont {Tan}},\ and\ \bibinfo {author}
  {\bibfnamefont {Y.}~\bibnamefont {Yu}},\ }\bibfield  {title} {\bibinfo
  {title} {Experimental realization of noncyclic geometric gates with shortcut
  to adiabaticity in a superconducting circuit},\ }\href@noop {} {\bibfield
  {journal} {\bibinfo  {journal} {Applied Physics Letters}\ }\textbf {\bibinfo
  {volume} {118}},\ \bibinfo {pages} {254002} (\bibinfo {year}
  {2021})}\BibitemShut {NoStop}%
\bibitem [{\citenamefont {Chen}\ \emph {et~al.}(2021)\citenamefont {Chen},
  \citenamefont {Qin}, \citenamefont {Wang}, \citenamefont {Miranowicz},\ and\
  \citenamefont {Nori}}]{chen2021shortcuts}%
  \BibitemOpen
  \bibfield  {author} {\bibinfo {author} {\bibfnamefont {Y.-H.}\ \bibnamefont
  {Chen}}, \bibinfo {author} {\bibfnamefont {W.}~\bibnamefont {Qin}}, \bibinfo
  {author} {\bibfnamefont {X.}~\bibnamefont {Wang}}, \bibinfo {author}
  {\bibfnamefont {A.}~\bibnamefont {Miranowicz}},\ and\ \bibinfo {author}
  {\bibfnamefont {F.}~\bibnamefont {Nori}},\ }\bibfield  {title} {\bibinfo
  {title} {Shortcuts to adiabaticity for the quantum rabi model: Efficient
  generation of giant entangled cat states via parametric amplification},\
  }\href@noop {} {\bibfield  {journal} {\bibinfo  {journal} {Physical Review
  Letters}\ }\textbf {\bibinfo {volume} {126}},\ \bibinfo {pages} {023602}
  (\bibinfo {year} {2021})}\BibitemShut {NoStop}%
\bibitem [{\citenamefont {Huang}\ \emph {et~al.}(2018)\citenamefont {Huang},
  \citenamefont {Kang}, \citenamefont {Chen}, \citenamefont {Shi},
  \citenamefont {Song},\ and\ \citenamefont {Xia}}]{huang2018quantum}%
  \BibitemOpen
  \bibfield  {author} {\bibinfo {author} {\bibfnamefont {B.-H.}\ \bibnamefont
  {Huang}}, \bibinfo {author} {\bibfnamefont {Y.-H.}\ \bibnamefont {Kang}},
  \bibinfo {author} {\bibfnamefont {Y.-H.}\ \bibnamefont {Chen}}, \bibinfo
  {author} {\bibfnamefont {Z.-C.}\ \bibnamefont {Shi}}, \bibinfo {author}
  {\bibfnamefont {J.}~\bibnamefont {Song}},\ and\ \bibinfo {author}
  {\bibfnamefont {Y.}~\bibnamefont {Xia}},\ }\bibfield  {title} {\bibinfo
  {title} {Quantum state transfer in spin chains via shortcuts to
  adiabaticity},\ }\href@noop {} {\bibfield  {journal} {\bibinfo  {journal}
  {Physical Review A}\ }\textbf {\bibinfo {volume} {97}},\ \bibinfo {pages}
  {012333} (\bibinfo {year} {2018})}\BibitemShut {NoStop}%
\bibitem [{\citenamefont {Mortensen}\ \emph {et~al.}(2018)\citenamefont
  {Mortensen}, \citenamefont {S{\o}rensen}, \citenamefont {M{\o}lmer},\ and\
  \citenamefont {Sherson}}]{mortensen2018fast}%
  \BibitemOpen
  \bibfield  {author} {\bibinfo {author} {\bibfnamefont {H.~L.}\ \bibnamefont
  {Mortensen}}, \bibinfo {author} {\bibfnamefont {J.~J.~W.}\ \bibnamefont
  {S{\o}rensen}}, \bibinfo {author} {\bibfnamefont {K.}~\bibnamefont
  {M{\o}lmer}},\ and\ \bibinfo {author} {\bibfnamefont {J.~F.}\ \bibnamefont
  {Sherson}},\ }\bibfield  {title} {\bibinfo {title} {Fast state transfer in a
  $\lambda$-system: a shortcut-to-adiabaticity approach to robust and resource
  optimized control},\ }\href@noop {} {\bibfield  {journal} {\bibinfo
  {journal} {New Journal of Physics}\ }\textbf {\bibinfo {volume} {20}},\
  \bibinfo {pages} {025009} (\bibinfo {year} {2018})}\BibitemShut {NoStop}%
\bibitem [{\citenamefont {Petiziol}\ \emph {et~al.}(2020)\citenamefont
  {Petiziol}, \citenamefont {Arimondo}, \citenamefont {Giannelli},
  \citenamefont {Mintert},\ and\ \citenamefont
  {Wimberger}}]{petiziol2020optimized}%
  \BibitemOpen
  \bibfield  {author} {\bibinfo {author} {\bibfnamefont {F.}~\bibnamefont
  {Petiziol}}, \bibinfo {author} {\bibfnamefont {E.}~\bibnamefont {Arimondo}},
  \bibinfo {author} {\bibfnamefont {L.}~\bibnamefont {Giannelli}}, \bibinfo
  {author} {\bibfnamefont {F.}~\bibnamefont {Mintert}},\ and\ \bibinfo {author}
  {\bibfnamefont {S.}~\bibnamefont {Wimberger}},\ }\bibfield  {title} {\bibinfo
  {title} {Optimized three-level quantum transfers based on frequency-modulated
  optical excitations},\ }\href@noop {} {\bibfield  {journal} {\bibinfo
  {journal} {Scientific reports}\ }\textbf {\bibinfo {volume} {10}},\ \bibinfo
  {pages} {1} (\bibinfo {year} {2020})}\BibitemShut {NoStop}%
\bibitem [{\citenamefont {Zhang}\ \emph {et~al.}(2021)\citenamefont {Zhang},
  \citenamefont {Yan}, \citenamefont {Lu},\ and\ \citenamefont
  {Feng}}]{zhang2021population}%
  \BibitemOpen
  \bibfield  {author} {\bibinfo {author} {\bibfnamefont {J.-L.}\ \bibnamefont
  {Zhang}}, \bibinfo {author} {\bibfnamefont {R.-Y.}\ \bibnamefont {Yan}},
  \bibinfo {author} {\bibfnamefont {X.-J.}\ \bibnamefont {Lu}},\ and\ \bibinfo
  {author} {\bibfnamefont {Z.-B.}\ \bibnamefont {Feng}},\ }\bibfield  {title}
  {\bibinfo {title} {Population transfer in a superconducting qutrit via
  shortcut to adiabaticity with optimized drivings},\ }\href@noop {} {\bibfield
   {journal} {\bibinfo  {journal} {Optics Communications}\ ,\ \bibinfo {pages}
  {127196}} (\bibinfo {year} {2021})}\BibitemShut {NoStop}%
\bibitem [{\citenamefont {Wang}\ and\ \citenamefont
  {Clerk}(2012)}]{wang2012using}%
  \BibitemOpen
  \bibfield  {author} {\bibinfo {author} {\bibfnamefont {Y.-D.}\ \bibnamefont
  {Wang}}\ and\ \bibinfo {author} {\bibfnamefont {A.~A.}\ \bibnamefont
  {Clerk}},\ }\bibfield  {title} {\bibinfo {title} {Using interference for high
  fidelity quantum state transfer in optomechanics},\ }\href@noop {} {\bibfield
   {journal} {\bibinfo  {journal} {Physical review letters}\ }\textbf {\bibinfo
  {volume} {108}},\ \bibinfo {pages} {153603} (\bibinfo {year}
  {2012})}\BibitemShut {NoStop}%
\bibitem [{\citenamefont {Tian}(2012)}]{tian2012adiabatic}%
  \BibitemOpen
  \bibfield  {author} {\bibinfo {author} {\bibfnamefont {L.}~\bibnamefont
  {Tian}},\ }\bibfield  {title} {\bibinfo {title} {Adiabatic state conversion
  and pulse transmission in optomechanical systems},\ }\href@noop {} {\bibfield
   {journal} {\bibinfo  {journal} {Physical review letters}\ }\textbf {\bibinfo
  {volume} {108}},\ \bibinfo {pages} {153604} (\bibinfo {year}
  {2012})}\BibitemShut {NoStop}%
\bibitem [{\citenamefont {Baksic}\ \emph {et~al.}(2017)\citenamefont {Baksic},
  \citenamefont {Belyansky}, \citenamefont {Ribeiro},\ and\ \citenamefont
  {Clerk}}]{baksic2017shortcuts}%
  \BibitemOpen
  \bibfield  {author} {\bibinfo {author} {\bibfnamefont {A.}~\bibnamefont
  {Baksic}}, \bibinfo {author} {\bibfnamefont {R.}~\bibnamefont {Belyansky}},
  \bibinfo {author} {\bibfnamefont {H.}~\bibnamefont {Ribeiro}},\ and\ \bibinfo
  {author} {\bibfnamefont {A.~A.}\ \bibnamefont {Clerk}},\ }\bibfield  {title}
  {\bibinfo {title} {Shortcuts to adiabaticity in the presence of a continuum:
  Applications to itinerant quantum state transfer},\ }\href@noop {} {\bibfield
   {journal} {\bibinfo  {journal} {Physical Review A}\ }\textbf {\bibinfo
  {volume} {96}},\ \bibinfo {pages} {021801} (\bibinfo {year}
  {2017})}\BibitemShut {NoStop}%
\bibitem [{\citenamefont {Aspelmeyer}\ \emph {et~al.}(2014)\citenamefont
  {Aspelmeyer}, \citenamefont {Kippenberg},\ and\ \citenamefont
  {Marquardt}}]{aspelmeyer2014cavity}%
  \BibitemOpen
  \bibfield  {author} {\bibinfo {author} {\bibfnamefont {M.}~\bibnamefont
  {Aspelmeyer}}, \bibinfo {author} {\bibfnamefont {T.~J.}\ \bibnamefont
  {Kippenberg}},\ and\ \bibinfo {author} {\bibfnamefont {F.}~\bibnamefont
  {Marquardt}},\ }\bibfield  {title} {\bibinfo {title} {Cavity optomechanics},\
  }\href@noop {} {\bibfield  {journal} {\bibinfo  {journal} {Reviews of Modern
  Physics}\ }\textbf {\bibinfo {volume} {86}},\ \bibinfo {pages} {1391}
  (\bibinfo {year} {2014})}\BibitemShut {NoStop}%
\bibitem [{\citenamefont {Yin}\ \emph {et~al.}(2015)\citenamefont {Yin},
  \citenamefont {Yang}, \citenamefont {Sun},\ and\ \citenamefont
  {Duan}}]{yin2015quantum}%
  \BibitemOpen
  \bibfield  {author} {\bibinfo {author} {\bibfnamefont {Z.-Q.}\ \bibnamefont
  {Yin}}, \bibinfo {author} {\bibfnamefont {W.}~\bibnamefont {Yang}}, \bibinfo
  {author} {\bibfnamefont {L.}~\bibnamefont {Sun}},\ and\ \bibinfo {author}
  {\bibfnamefont {L.}~\bibnamefont {Duan}},\ }\bibfield  {title} {\bibinfo
  {title} {Quantum network of superconducting qubits through an optomechanical
  interface},\ }\href@noop {} {\bibfield  {journal} {\bibinfo  {journal}
  {Physical Review A}\ }\textbf {\bibinfo {volume} {91}},\ \bibinfo {pages}
  {012333} (\bibinfo {year} {2015})}\BibitemShut {NoStop}%
\bibitem [{\citenamefont {Vasilev}\ \emph {et~al.}(2009)\citenamefont
  {Vasilev}, \citenamefont {Kuhn},\ and\ \citenamefont
  {Vitanov}}]{vasilev2009optimum}%
  \BibitemOpen
  \bibfield  {author} {\bibinfo {author} {\bibfnamefont {G.}~\bibnamefont
  {Vasilev}}, \bibinfo {author} {\bibfnamefont {A.}~\bibnamefont {Kuhn}},\ and\
  \bibinfo {author} {\bibfnamefont {N.}~\bibnamefont {Vitanov}},\ }\bibfield
  {title} {\bibinfo {title} {Optimum pulse shapes for stimulated raman
  adiabatic passage},\ }\href@noop {} {\bibfield  {journal} {\bibinfo
  {journal} {Physical Review A}\ }\textbf {\bibinfo {volume} {80}},\ \bibinfo
  {pages} {013417} (\bibinfo {year} {2009})}\BibitemShut {NoStop}%
\bibitem [{\citenamefont {Tittel}\ \emph {et~al.}(1998)\citenamefont {Tittel},
  \citenamefont {Brendel}, \citenamefont {Gisin}, \citenamefont {Herzog},
  \citenamefont {Zbinden},\ and\ \citenamefont
  {Gisin}}]{tittel1998experimental}%
  \BibitemOpen
  \bibfield  {author} {\bibinfo {author} {\bibfnamefont {W.}~\bibnamefont
  {Tittel}}, \bibinfo {author} {\bibfnamefont {J.}~\bibnamefont {Brendel}},
  \bibinfo {author} {\bibfnamefont {B.}~\bibnamefont {Gisin}}, \bibinfo
  {author} {\bibfnamefont {T.}~\bibnamefont {Herzog}}, \bibinfo {author}
  {\bibfnamefont {H.}~\bibnamefont {Zbinden}},\ and\ \bibinfo {author}
  {\bibfnamefont {N.}~\bibnamefont {Gisin}},\ }\bibfield  {title} {\bibinfo
  {title} {Experimental demonstration of quantum correlations over more than 10
  km},\ }\href@noop {} {\bibfield  {journal} {\bibinfo  {journal} {Physical
  Review A}\ }\textbf {\bibinfo {volume} {57}},\ \bibinfo {pages} {3229}
  (\bibinfo {year} {1998})}\BibitemShut {NoStop}%
\bibitem [{\citenamefont {Teufel}\ \emph {et~al.}(2011)\citenamefont {Teufel},
  \citenamefont {Donner}, \citenamefont {Li}, \citenamefont {Harlow},
  \citenamefont {Allman}, \citenamefont {Cicak}, \citenamefont {Sirois},
  \citenamefont {Whittaker}, \citenamefont {Lehnert},\ and\ \citenamefont
  {Simmonds}}]{teufel2011sideband}%
  \BibitemOpen
  \bibfield  {author} {\bibinfo {author} {\bibfnamefont {J.~D.}\ \bibnamefont
  {Teufel}}, \bibinfo {author} {\bibfnamefont {T.}~\bibnamefont {Donner}},
  \bibinfo {author} {\bibfnamefont {D.}~\bibnamefont {Li}}, \bibinfo {author}
  {\bibfnamefont {J.~W.}\ \bibnamefont {Harlow}}, \bibinfo {author}
  {\bibfnamefont {M.}~\bibnamefont {Allman}}, \bibinfo {author} {\bibfnamefont
  {K.}~\bibnamefont {Cicak}}, \bibinfo {author} {\bibfnamefont {A.~J.}\
  \bibnamefont {Sirois}}, \bibinfo {author} {\bibfnamefont {J.~D.}\
  \bibnamefont {Whittaker}}, \bibinfo {author} {\bibfnamefont {K.~W.}\
  \bibnamefont {Lehnert}},\ and\ \bibinfo {author} {\bibfnamefont {R.~W.}\
  \bibnamefont {Simmonds}},\ }\bibfield  {title} {\bibinfo {title} {Sideband
  cooling of micromechanical motion to the quantum ground state},\ }\href@noop
  {} {\bibfield  {journal} {\bibinfo  {journal} {Nature}\ }\textbf {\bibinfo
  {volume} {475}},\ \bibinfo {pages} {359} (\bibinfo {year}
  {2011})}\BibitemShut {NoStop}%
\bibitem [{\citenamefont {Chan}\ \emph {et~al.}(2011)\citenamefont {Chan},
  \citenamefont {Alegre}, \citenamefont {Safavi-Naeini}, \citenamefont {Hill},
  \citenamefont {Krause}, \citenamefont {Gr{\"o}blacher}, \citenamefont
  {Aspelmeyer},\ and\ \citenamefont {Painter}}]{chan2011laser}%
  \BibitemOpen
  \bibfield  {author} {\bibinfo {author} {\bibfnamefont {J.}~\bibnamefont
  {Chan}}, \bibinfo {author} {\bibfnamefont {T.~M.}\ \bibnamefont {Alegre}},
  \bibinfo {author} {\bibfnamefont {A.~H.}\ \bibnamefont {Safavi-Naeini}},
  \bibinfo {author} {\bibfnamefont {J.~T.}\ \bibnamefont {Hill}}, \bibinfo
  {author} {\bibfnamefont {A.}~\bibnamefont {Krause}}, \bibinfo {author}
  {\bibfnamefont {S.}~\bibnamefont {Gr{\"o}blacher}}, \bibinfo {author}
  {\bibfnamefont {M.}~\bibnamefont {Aspelmeyer}},\ and\ \bibinfo {author}
  {\bibfnamefont {O.}~\bibnamefont {Painter}},\ }\bibfield  {title} {\bibinfo
  {title} {Laser cooling of a nanomechanical oscillator into its quantum ground
  state},\ }\href@noop {} {\bibfield  {journal} {\bibinfo  {journal} {Nature}\
  }\textbf {\bibinfo {volume} {478}},\ \bibinfo {pages} {89} (\bibinfo {year}
  {2011})}\BibitemShut {NoStop}%
\bibitem [{\citenamefont {Pirkkalainen}\ \emph {et~al.}(2013)\citenamefont
  {Pirkkalainen}, \citenamefont {Cho}, \citenamefont {Li}, \citenamefont
  {Paraoanu}, \citenamefont {Hakonen},\ and\ \citenamefont
  {Sillanp{\"a}{\"a}}}]{pirkkalainen2013hybrid}%
  \BibitemOpen
  \bibfield  {author} {\bibinfo {author} {\bibfnamefont {J.-M.}\ \bibnamefont
  {Pirkkalainen}}, \bibinfo {author} {\bibfnamefont {S.}~\bibnamefont {Cho}},
  \bibinfo {author} {\bibfnamefont {J.}~\bibnamefont {Li}}, \bibinfo {author}
  {\bibfnamefont {G.}~\bibnamefont {Paraoanu}}, \bibinfo {author}
  {\bibfnamefont {P.}~\bibnamefont {Hakonen}},\ and\ \bibinfo {author}
  {\bibfnamefont {M.}~\bibnamefont {Sillanp{\"a}{\"a}}},\ }\bibfield  {title}
  {\bibinfo {title} {Hybrid circuit cavity quantum electrodynamics with a
  micromechanical resonator},\ }\href@noop {} {\bibfield  {journal} {\bibinfo
  {journal} {Nature}\ }\textbf {\bibinfo {volume} {494}},\ \bibinfo {pages}
  {211} (\bibinfo {year} {2013})}\BibitemShut {NoStop}%
\end{thebibliography}%
	
\end{document}